\documentclass[a4paper,11pt]{article}
\pdfoutput=1 

\usepackage{jheppub} 
\usepackage[dvipsnames]{xcolor}
\usepackage{graphicx,color}
\usepackage{hyperref}
\usepackage{caption}
\usepackage{subcaption}
\usepackage{amsmath,amssymb}

\newcommand{\af}[1]{\textcolor{teal}{#1}}

\renewcommand{\O}{\mathcal{O}}

\arxivnumber{} 

\title{Toward an Effective Theory of the Volume Modulus}

\author[a]{Naman Agarwal,}
\author[a,b]{Andrew R. Frey,}
\author[c]{and Bret Underwood}

\affiliation[a]{Department of Physics \& Astronomy, University of Manitoba\\
Winnipeg, Manitoba R3T 2N2, Canada}
\affiliation[b]{Department of Physics, University of Winnipeg\\ Winnipeg, Manitoba R3B 2E9, Canada}
\affiliation[c]{Department of Physics, Pacific Lutheran University\\ Tacoma, WA 98447, USA}

\emailAdd{agarwaln@myumanitoba.ca}
\emailAdd{a.frey@uwinnipeg.ca}
\emailAdd{bret.underwood@plu.edu}

\abstract{
We investigate the 4-dimensional effective theory of the warped volume modulus in the presence of stabilizing effects from gaugino condensation by analyzing the linearized 10-dimensional supergravity equations of motion.
Warping is generally expected to scale down the masses of bulk modes to the IR scale at the tip of a throat.
We 
find that
the mass of the warped volume modulus evades expectations and is largely insensitive to the effects of warping, even in strongly warped backgrounds. Instead, the mass is parametrically tied to the 4-dimensional AdS curvature scale $m^2 \sim {\mathcal O}(1) |\hat R_{\rm AdS}|$, presenting a challenge for scale separation in these backgrounds.
We trace this effect to a universal contribution arising from the 10-dimensional equations of motion, and comment on the importance of a 10-dimensional treatment of the warped volume modulus for effective field theories and model building.

}


\begin{document}
\maketitle
\flushbottom

\section{Introduction}\label{sec:intro}


In order to analyze the vacua and stability of compactifications from string theory, it is helpful to construct a 4-dimensional effective field theory of the light fields of the compactification, often called moduli.
The moduli arise as fluctuations of the 10-dimensional fields, such as the metric, $p$-form gauge fields, or D-brane degrees of freedom, and their effective theory can be constructed by reducing the higher-dimensional equations of motion to 4-dimensions.
Ingredients introduced in the higher-dimensional theory, such as fluxes of $p$-form fields wrapping cycles of the internal manifold, orientifold planes, and D-branes, show up as contributions to the effective potential as functions of the moduli, determining the vacua of the theory and the stabilization of moduli as massive fluctuations about the vacua \cite{GKP,KKLT}.
However, these ingredients necessary for the construction of the vacua and for moduli stabilization are energy sources in the higher-dimensional theory, and as a result cause warping and source background profiles of the 10-dimensional fields. 

In the presence of metric warping and background profiles, fluctuations of 10-dimensional fields behave differently than their unwarped counterparts.
For example, many of the fluctuations of the 10-dimensional fields mix kinematically through the background and warping into new fluctuations \cite{Frey:2002hf,Giddings_Maharana,STUD,FTUD,BreathingMode,1308.0323,Cownden:2016hpf,Frey:2025rvf}.
A prototypical example of this mixing between 10-dimensional fields into new fluctuations is the universal volume modulus (reviewed in more detail later in this section).
In the absence of warping, the universal volume modulus is identified as a fluctuation of the overall scale of the internal metric, accompanied by a Weyl rescaling of the 4-dimensional metric to ensure the dimensionally reduced theory is in 4-dimensional Einstein frame.
However, when the background is warped as in \cite{GKP}, fluctuations of the overall scale of the internal metric do not correspond to an independent degree of freedom.
Instead, the warped volume modulus arises as a mixture of fluctuations of 10-dimensional fields, including the warp factor, Weyl rescaling factor, 4-form gauge potential, and kinematic ``compensator" fluctuations of the metric \cite{FTUD}.
Since the warped volume modulus corresponds to fluctuations of a different set of 10-dimensional fields compared to the unwarped volume modulus, this difference can affect the functional dependence of the resulting effective theory on the warped volume modulus.
As another example, K\"ahler moduli and axions descending from internal metric deformations and $p$-form fields, respectively, mix with each other through the moduli space metric in the presence of warping and background fluxes \cite{1308.0323,Frey:2025rvf}.

Another effect of warped backgrounds is 
a general expectation for wavefunctions and mass scales of 10-dimensional fluctuations 
to be pulled down by strongly warped throats.
For example, the wavefunctions of KK gravitons in strongly warped throats are localized to deep in the IR end of the throat compared to the zero mode, with masses scaled down by corresponding exponentially small warping factors
\cite{Randall:1999ee,DeWolfeGiddings,Kofman:2005yz,Firouzjahi:2005qs,ValeixoBento:2022qca}.
Fluctuations of other 10-dimensional fields, such as massive bulk scalar fields \cite{Burgess:2006mn} and the dilaton \cite{Frey:2006wv} show similar behavior in which wavefunctions localize towards the tip of the throat and mass scales become warped down.
Similarly, classically stabilized complex structure moduli are believed to experience similar effects \cite{DT,Bena:2018fqc,Blumenhagen:2019qcg,Bena:2019sxm,Randall:2019ent,Dudas:2019pls,Lust:2022xoq}.
While a massless volume modulus fluctuation should have a constant wavefunction throughout a strongly warped background, a naive expectation based on the prior work cited above might be that a massive warped volume modulus would be affected by warping in similar ways as other bulk fluctuations.

One of the best-studied models of compactification with fluxes and warped backgrounds is the KKLT model \cite{KKLT} where bulk $p$-form fluxes, D-branes, orientifold planes, and gaugino condensation on D7-branes stabilize the volume modulus of the compactification with a negative (AdS) vacuum energy.
However, a puzzle arises because these stabilizing ingredients generate significant metric warping, including strongly warped throats, the effects of which are not apparent in the simple KKLT effective theory.
As discussed above, from a 10-dimensional perspective warping should have profound effects: it should redefine the volume modulus as a mixture of metric and warp factor fluctuations, affecting the resulting effective theory, and 
previous work on 10-dimensional fluctuations suggest that
the modulus mass and wavefunction 
should be pulled down to the warped IR scale of a throat. These effects stand in contrast to the KKLT effective theory \cite{KKLT}, where the volume modulus is treated as a simple rescaling of the metric and its mass is controlled primarily by the AdS vacuum energy scale.
The tension between these two perspectives motivates a 10-dimensional analysis.
In recent years, there has been considerable progress in understanding the 10-dimensional origin of gaugino condensation on D7-branes and its effect on other 10-dimensional fields, both from the 10-dimensional equations of motion with gaugino sources \cite{Baumann:2008kq, Baumann:2010sx,Moritz:2017xto,Gautason:2018gln,Hamada:2018qef,Carta:2019rhx,Kallosh:2019axr,Kallosh:2019oxv,Hamada:2019ack,Gautason:2019jwq,Kachru:2019dvo,Hamada:2021ryq}
and from 10-dimensional supersymmetry variations modified to account for gaugino condensation \cite{Koerber:2007xk,Koerber:2008sx,Bena:2019mte,Grana:2020hyu,Grana:2022nyp}.
However, most previous studies have used the unwarped volume modulus in attempts to understand the resulting effective theory.
Note that the value of the gaugino condensate itself may depend on the value of the warp factor along the D7-branes, especially if the branes extend into a strongly warped region; however, the warp factor and gaugino condensate (along with other bulk fields) are determined self-consistently by the equations of motion for the background. 
Another important effect, which we include in our model, is how the gaugino condensate fluctuates along with the modulus.
In this article, we make progress towards a 10-dimensional analysis of the warped volume modulus in backgrounds with gaugino condensation, with a focus on understanding how the profile and mass of the volume modulus is affected by strongly warped backgrounds.


In the rest of this section, we review the massless warped volume modulus in the backgrounds of \cite{GKP}. We demonstrate why the usual unwarped volume modulus fails as a 10-dimensional fluctuation in the presence of warping and identify the correct warped fluctuation as a mixture of fluctuations of the metric and warp factors. We then show how the difference between the unwarped and warped volume modulus can show up in the resulting effective potential.
In Section \ref{sec:eom} we briefly review the background 10-dimensional fields sourced by the gaugino condensation of \cite{KKLT}, and set up the linearized equations of motion for fluctuations of the 10-dimensional fields in this background corresponding to the warped volume modulus.
In Section \ref{sec:Simplified Toy Models}, we solve the linearized equations of motion for some simplified cases, finding that the 4-dimensional mass of the corresponding volume modulus is typically $m^2 \sim {\mathcal O}(1) |\hat R_{AdS}|$, where $\hat R_{AdS}$ is the 4-dimensional anti de-Sitter vacuum energy.
Interestingly, we find that this result holds even in strongly warped backgrounds, so that the mass of the warped volume modulus is not warped down to the IR scale of a strongly warped throat. 
Instead, the warped volume modulus is generally insensitive to the presence of warping, and there is no clear parametric separation between the mass and the AdS curvature scale in general.
We conclude with a discussion in Section \ref{sec:discuss} on the comparison of our results to those found using the effective theory of \cite{KKLT}, general lessons learned, and some possible future directions.

Before we begin, a word on notation:
In the following sections (including the appendices) we will carry out a dual perturbative expansion, one an expansion in the strength of the gaugino condensate (labeled as $\epsilon$), and the second an expansion in fluctuations around the background.
For ease and clarity of reading, we use the following notation. A variable $\Phi(x,y)$ is given at all orders in the gaugino condensate and fluctuations (a variable may be suppressed if the field does not depend on it; for example, the background does not depend on $x$ and may be denoted $\Phi(y)$).
A numerical subscript indicates the order in the gaugino condensate, i.e.~$0$ represents the classical background and fluctuations ($\mathcal{O}(\epsilon^0)$). The first nontrivial corrections due to the gaugino are first or second order depending on the field ($\Phi_1(y)$ or $\Phi_2(y)$ in the background). 
Fluctuations will all have spacetime dependence $c(x)$, and the dependence on the internal space ($y$ directions) will be indicated by adding a $\delta$ to the corresponding background variable, so the fluctuation in field $\Phi$ is $c(x)\delta\Phi(y)$ (that is, $\Phi(x,y)=\Phi(y)+c(x)\delta\Phi(y)$). 
A numerical subscript in parentheses indicates the rank of a form, so $\Phi_{(1)}$ is a vector field.

\subsection{The Massless Warped Volume Modulus}
\label{sec:WarpedVolModulus}

It is helpful to first review how the universal volume modulus arises as a massless fluctuation of 10-dimensional fields in the type IIB flux background of \cite{GKP}, following \cite{FTUD}, before including non-perturbative effects.\footnote{We therefore omit the subscript $0$ in this section because all fields are classical and do not depend on the gaugino condensate.}
We will also show how the common practice of identifying the universal volume modulus as an overall scaling of the internal metric is not a consistent fluctuation in 10-dimensions, and can lead to incorrect results for the moduli-dependence of potential energy terms in the lower-dimensional effective field theory.

The background in type IIB supergravity \cite{GKP} is given by a warped product metric on a Ricci-flat internal space
\begin{equation}
    ds^2 = e^{2A(y)}\hat \eta_{\mu\nu} dx^\mu dx^\nu + e^{-2A(y)} \tilde{g}_{mn} dy^m dy^n\, , \\
    \label{eq:GKP_metric}
\end{equation}
as well as fluxes\footnote{The quantities $\hat \epsilon, \tilde\star$ are with respect to the $\hat\eta_{\mu\nu}, \tilde g_{mn}$ metrics, respectively, and $\tilde d = dy^m \partial_m $ is an exterior derivative with respect to the internal coordinates.}
\begin{align}
    & G_{(3)}^+\,, \hspace{.3in} \tilde F_{(5)} = \hat \epsilon \wedge \tilde d e^{4A(y)} + \tilde\star\, \tilde d e^{-4A(y)}\, ,
    \label{eq:GKP_fluxes}
\end{align}
where $G_{(3)} = F_{(3)} - \tau H_{(3)}$ wraps 3-cycles in the internal space, $\tilde F_{(5)}$ is self-dual $\star \tilde F_{(5)} = \tilde F_{(5)}$, and $\tau$ is the complex axio-dilaton $\tau = C_{(0)} + i e^{-\phi}$.
The $G_{(3)}^+$ flux is imaginary self-dual (ISD)
\begin{align}
    &\tilde\star G_{(3)}^\pm = \pm i G_{(3)}^\pm
\end{align}
(flux $G_{(3)}^-$ is imaginary anti-self-dual (IASD)).
Including effective D3-brane sources (this may include D3 and O3 sources, as well as D7 and O7 sources in the orientifold limit), the background warp factor is determined by the expression
\begin{equation}
    -\tilde \nabla^2 e^{-4A} = \frac{G_{abc}\bar{G}^{\widetilde{abc}}}{12\, {\rm Im}\tau} + 2 \kappa_{10}^2 T_3 \tilde \rho_3^{\rm loc}\, .
    \label{eq:GKP_Poisson}
\end{equation}

In unwarped backgrounds, the universal volume modulus usually corresponds to an overall ``scaling mode'' of the internal space $\tilde g_{mn} \rightarrow L^2(x) \tilde g_{mn}$ (along with a corresponding Weyl-rescaling of the 4-dimensional metric to preserve 4-dimensional Einstein frame). 
Extending this expectation to the warped background (\ref{eq:GKP_metric}), a ``scaling mode'' proposal for the volume modulus takes the form \cite{GKP,DeWolfeGiddings}
\begin{equation}
ds^2 = e^{2A(y)} L(x)^{-6}\, \hat \eta_{\mu\nu} dx^\mu dx^\nu + e^{-2A(y)} L(x)^2\, \tilde g_{mn} dy^m dy^n\, .
\label{eq:unwarpedVolModulus}
\end{equation}
Consider an effective theory constructed by dimensional reduction of the 10-dimensional metric with volume modulus fluctuations given by (\ref{eq:unwarpedVolModulus}) as in \cite{DeWolfeGiddings,Giddings:2003zw,Hertzberg:2007wc,SimpledS,Haque:2008jz,Danielsson:2009ff,Wrase:2010ew,VanRiet:2011yc,Kachru:2019dvo}.
The 4-dimensional scalar potential arises by dimensional reduction of several terms, including 3-form flux, 5-form flux, metric warping, and local sources
\begin{equation}
    V_4 = V_{\rm flux} + V_{\rm warp} + V_{\rm loc}\, .
    \label{eq:4dEFTPotential}
\end{equation}
For example, 
the contribution to the scalar potential from 
an anti D3-brane with the scaling ansatz (\ref{eq:unwarpedVolModulus}) becomes
\begin{equation}
    V_{\rm D3} = \frac{2 T_3 e^{4A(y^*)}}{L^{12}} = \frac{2 T_3 e^{4A(y^*)}}{\left({\rm Im}\,\rho\right)^3}\, ,
    \label{eq:antiD3unwarpedVol}
\end{equation}
where it is common to identify the imaginary part of the complex K\"ahler modulus as ${\rm Im}\, \rho = L^4$ and $A(y_*)$ is the warp factor evaluated at the location of the anti D3-brane in the internal space $\{y_m\} = \{y_m^*\}$.

Unfortunately, the proposal of (\ref{eq:unwarpedVolModulus}) as the 10-dimensional description of a volume modulus fails in flux compactifications for several reasons.
First, notice that the background equation determining the warp factor (\ref{eq:GKP_Poisson}) is not invariant under the scaling (\ref{eq:unwarpedVolModulus}), thus scaling by $L$ is not a zero mode of the background, even in the zero-momentum limit in which the scaling is taken to be a constant. Since the volume modulus is expected to be massless in the 4-dimensional effective theory corresponding to this background \cite{GKP,KKLT}, the scaling (\ref{eq:unwarpedVolModulus}) cannot correspond to the volume modulus in this background.
Second, the scaling (\ref{eq:unwarpedVolModulus}) also fails to satisfy the 10-dimensional Einstein equations as needed in order to correspond to an independent degree of freedom in a lower-dimensional effective theory. 
In particular, the off-diagonal Einstein equations with the metric (\ref{eq:unwarpedVolModulus}) serve as a kinematic constraint \cite{DT,FTUD} on the fluctuations,
\begin{align}
\label{eq:unwarpedMuM}
G_{\mu m} &= 4\, \partial_\mu \left(L^2(x)\right) \partial_m A(y) = 0\, .
\end{align}
Unless the background is unwarped $\partial_m A(y) = 0$ (in contradiction to (\ref{eq:GKP_Poisson})), the scaling (\ref{eq:unwarpedVolModulus}) cannot be a solution, and thus cannot be a kinematic fluctuation of the metric.
It is tempting to consider the failure of the scaling ansatz (\ref{eq:unwarpedVolModulus}) to satisfy (\ref{eq:unwarpedMuM}) as 
an approximation for weakly warped backgrounds $\partial_m A \approx 0$.
However, the weakly warped condition is impossible to satisfy in general in the background (\ref{eq:GKP_Poisson}) (particularly near local sources such as D3 and O3-planes), and strongly warped regions violating (\ref{eq:unwarpedMuM}) are necessary for the construction of de Sitter vacua of \cite{KKLT}.
The failure of the scaling ansatz (\ref{eq:unwarpedVolModulus}) can be seen as arising from an incomplete 
identification of the metric degrees of freedom for scalar fluctuations, with parallels to cosmological perturbation theory \cite{BreathingMode}.
The effects of accurately identifying the volume modulus in 10-dimensions are important for 4-dimensional effective field theory and model building. The failure of the scaling ansatz (\ref{eq:unwarpedVolModulus}) 
to correspond to a valid volume modulus leads to an incorrect functional dependence of the effective theory and effective potential (\ref{eq:antiD3unwarpedVol}) on the volume modulus.

Instead of (\ref{eq:unwarpedVolModulus}), in the warped flux backgrounds of (\ref{eq:GKP_metric}) the volume modulus arises as a dynamical shift of the warp factor \cite{Giddings_Maharana,FTUD}
\begin{equation}
    e^{-4A(y)}\rightarrow e^{-4A(x,y)} = e^{-4A(y)} + c(x)
    \label{eq:WarpedVolModulusShift}
\end{equation}
with $\tilde g_{mn}$ unchanged.
Notice that $c(x)$ is manifestly a zero mode of (\ref{eq:GKP_Poisson}) as desired, in contrast to the simple scaling of the internal metric.
Together with a Weyl-rescaling and a 10-dimensional metric fluctuation $\delta g_{\mu m}$ known as a ``compensator'', the warped volume modulus corresponds to the metric deformation \cite{FTUD}
\begin{align}
ds^2 = e^{2\Omega(x)} e^{2 A(x,y)} \hat \eta_{\mu\nu} + 2 e^{2 A(y)} \left(\partial_\mu c\right)  \delta B_m(y)\, dx^\mu dy^m + e^{-2A(x,y)} \tilde g_{mn} dy^m dy^n
\label{eq:WarpedVolModulus}
\end{align}
where the volume modulus-dependent Weyl-factor is
\begin{equation}
    e^{-2\Omega(x)} = \frac{\int d^6y \sqrt{\tilde g}\, e^{-4A(x,y)}}{\int d^6y \sqrt{\tilde g}} = \frac{\int d^6y \sqrt{\tilde g}\, e^{-4A(y)}}{\int d^6y \sqrt{\tilde g}} + c(x) = \frac{V_W}{\tilde V_{CY}} + c(x)\, ,
    \label{eq:GKPWeyl}
\end{equation}
and we defined the warped and unwarped volumes as $V_W = \int d^6y \sqrt{\tilde g}\, e^{-4A}$, $\tilde V_{CY} = \int d^6y \sqrt{\tilde g}$, respectively.
The compensator wavefunction takes the form
\begin{equation}
   \delta B_m(y) = \partial_m \delta K(y)
\end{equation}
where $\delta K(y)$ solves the equation
\begin{equation} \label{KCompensator}
    \tilde \nabla^2 \delta K = \frac{V_W}{\tilde V_{CY}} - e^{-4A(y)}\, .
\end{equation}
The metric\footnote{Along with a corresponding compensator in $\tilde F_{(5)}$.} (\ref{eq:WarpedVolModulus}) is a zero-mode of the background (\ref{eq:GKP_metric}-\ref{eq:GKP_Poisson}), solves the constraints arising from the 10-dimensional Einstein equations, and reduces to the correct form upon dimensional reduction \cite{FTUD}, so it is the correct form of the warped volume modulus.
Furthermore, fluctuations of (\ref{eq:WarpedVolModulus}) combine into gauge-invariant combinations with respect to 10-dimensional diffeomorphisms so that the resulting dimensional reduction is well-defined \cite{BreathingMode}.
Notice that in the limit of everywhere-weak warping $e^{-4A(y)}\approx {\rm const}$, the metric (\ref{eq:WarpedVolModulus}) \emph{approximately} takes the form of a scaling ansatz (\ref{eq:unwarpedVolModulus}) (up to a shift in the definition of the modulus).
However, in order to be a true zero mode of the background and satisfy the constraints, the volume modulus must fundamentally arise as the shift (\ref{eq:WarpedVolModulusShift}).

When we use the correct identification of the volume modulus from (\ref{eq:WarpedVolModulus}), the uplifting contribution to the effective potential from an anti D3-brane (\ref{eq:antiD3unwarpedVol}) becomes instead
\begin{align}
    V_{D3} &= 2T_3 e^{4A(x,y^*)} e^{4\Omega} = \frac{2 T_3}{\left({\rm Im}\,\rho+e^{-4A(y^*)}-V_W/\tilde V_{CY}\right)} \frac{1}{({\rm Im}\,\rho)^2}\, ,
    \label{eq:antiD3WarpedVol}
\end{align}
where ${\rm Im}\, \rho = e^{-2\Omega} = V_W/\tilde V_{CY} + c(x)$ is now correctly identified as the K\"ahler modulus\footnote{The axion associated with the 4-form gauge potential $C_4$ becomes the corresponding real part, although the dimensional reduction is non-trivial and requires flux as well as metric compensators \cite{1308.0323}. 
} \cite{FTUD}.
The weakly warped, large volume limit of the metric (\ref{eq:WarpedVolModulus}) is found as the limit of a large constant warp factor $e^{-4A(y)}\approx V_W/\tilde V_{CY} \gg 1$, which sets the background volume, with the mode $c(x)$ in (\ref{eq:WarpedVolModulusShift}) representing a small fluctuation. 
In this unwarped limit, the anti D3-brane potential becomes
\begin{equation}
    V_{D3} \approx \frac{2T_3}{\left({\rm Im}\,\rho\right)^3}\, ,
\end{equation}
which matches the ``scaling mode'' result (\ref{eq:antiD3unwarpedVol}) (after taking into account that the weakly warped, large volume limit for the scaling metric (\ref{eq:unwarpedVolModulus}) is $e^{-4A(y)}\approx 1$).
In a strongly warped background, however, the potential (\ref{eq:antiD3WarpedVol}) has different behaviors.
For example, when the anti D3-brane is located in a strongly warped region where $e^{-4A(y^*)} \gg V_W/\tilde V_{CY}$, we see a different functional dependence naturally emerge\footnote{
It has been argued \cite{KKLMMT} that, at the tip of a strongly warped throat, the warp factor $A(y)$ scales with an additional power of ${\rm Im}\, \rho$ (though it is unclear how this works in the case of a spacetime-dependent fluctuation $\rho(x)$, or somewhere other than the tip of a warped throat). This changes the proposed functional dependence using the scaling ansatz of the contribution to the potential from an anti D3-brane (\ref{eq:antiD3unwarpedVol}) at the tip of a throat to be instead
$V_{\rm D3} = \left({\rm Im}\,\rho\right)^{-2}$, in agreement with (\ref{eq:antiD3WarpedVolTip}). We see that this line of argument was only needed because the incorrect scaling ansatz was used for the volume modulus.} compared to (\ref{eq:antiD3unwarpedVol})
\begin{equation}
    V_{D3} \approx \frac{2T_3 e^{4A(y^*)}}{({\rm Im}\,\rho)^2}\, .
    \label{eq:antiD3WarpedVolTip}
\end{equation}
Alternatively, consider an internal space with strongly warped regions with $e^{-4A(y)} \gg 1$ so that $V_W/\tilde V_{CY} \gg 1$, but with the anti D3-brane located in a weakly warped region $e^{-4A(y*)} \approx 1$. 
In this scenario, the potential becomes
\begin{equation}
    V_{D3} \approx \frac{2T_3}{\left({\rm Im}\, \rho - V_W/\tilde V_{CY}\right)\left({\rm Im}\,\rho\right)^2}\, ,
    \label{eq:antiD3warpedVolLimit2}
\end{equation}
which is modified compared to the ``scaling mode" result (\ref{eq:antiD3unwarpedVol}) due to the shift in the denominator of (\ref{eq:antiD3warpedVolLimit2}).
While we have focused on the contribution from an anti D3-brane, similar differences in the functional dependence arise for other terms in the effective potential (\ref{eq:4dEFTPotential}) in  strongly warped backgrounds.
Thus it is essential to start with the correct identification of the 10-dimensional fluctuations corresponding to the volume modulus, otherwise one will end up with an incorrect 4-dimensional effective potential with the wrong functional dependence.

As one final perspective on the challenges of the scaling ansatz (\ref{eq:unwarpedVolModulus}), note that a scaling of the form $\tilde g_{mn} \rightarrow e^{2u}\tilde g_{mn}$ can be compensated by a scaling of the warp factor $e^{2A}\rightarrow e^{2u} e^{2A}$, leaving the internal space invariant. This transfers the factor $e^{2u}$ into a rescaling of the 4-dimensional space only, which is removed by a Weyl rescaling to 4-dimensional Einstein frame \cite{Giddings_Maharana,FTUD}, indicating the rescaling is a gauge redundancy.
Indeed, the most general 4-dimensional scalar perturbations of the background metric
\begin{align}
    ds_{10}^2 =& e^{2A(y)}\left[(1-2\psi(x,y))\hat g_{\mu\nu} + 2 \hat \nabla_\mu \partial_\nu K(x,y)\right]dx^\mu dx^\nu + e^{2A(y)} \partial_\mu B_m(x,y)\ dx^\mu dy^m \nonumber \\
    & e^{-2A(y)} \left(\tilde g_{mn}+2\phi_{mn}(x,y)\right)dy^m dy^n\, ,
\end{align}
organize themselves into combinations that are invariant under 10-dimensional diffeomorphisms \cite{BreathingMode}
\begin{align}
    \Phi_{mn} &= \phi_{mn} + e^{4A(y)} \left(\partial^{\tilde p} A\right)\left(B_p - \partial_p K\right) \tilde g_{mn} - \tilde \nabla_{(m} \left[e^{4A(y)} \left(B_{n)} - \partial_{n)} K\right)\right]\,; \label{eq:gaugeInvPhi}\\
    \Psi &= \psi + e^{4A(y)} \left(\partial^{\tilde p} A\right) \left(B_p - \partial_p K\right)\,. \label{eq:gaugeInvPsi}
\end{align}
Even if we restrict to a diagonal internal metric perturbation $\phi_{mn}(x,y) = \phi(x,y) \tilde g_{mn}$, the gauge-invariant fluctuation (\ref{eq:gaugeInvPhi}) is not diagonal $\Phi_{mn} \neq \Phi \tilde g_{mn}$ in the presence of warped backgrounds due to the derivatives of the warp factor and compensators in the last term. Thus, identifying the warped volume modulus as a simple scaling of the internal metric as in (\ref{eq:unwarpedVolModulus}) is not an allowed gauge-invariant degree of freedom.

So far this analysis has considered only a massless volume modulus, as a zero mode of the background (\ref{eq:GKP_metric}),(\ref{eq:GKP_fluxes}),(\ref{eq:GKP_Poisson}).
In the next section, following \cite{KKLT}, we will model the effect of stabilizing the volume modulus with non-perturbative effects due to gaugino condensation on D7-branes.

\section{Equations of Motion}\label{sec:eom}

Given the potential importance of KKLT models for de Sitter compactifications in string theory, there has been a significant effort to understand the effects of gaugino condensation at a 10-dimensional level. There are two basic approaches. 
One is to consider the D7-brane gaugino condensate as a source for the classical 10-dimensional supergravity equations of motion, taking into account the couplings of bulk fields to the brane \cite{Baumann:2008kq, Baumann:2010sx,Moritz:2017xto,Gautason:2018gln,Hamada:2018qef,Carta:2019rhx,Kallosh:2019axr,Kallosh:2019oxv,Hamada:2019ack,Gautason:2019jwq,Kachru:2019dvo,Hamada:2021ryq}.
The second method is to solve the 10-dimensional supersymmetry variations modified to account for gaugino condensation, as in \cite{Koerber:2007xk,Koerber:2008sx,Bena:2019mte,Grana:2020hyu,Grana:2022nyp}. 
We follow the first approach. We first present the equations of motion that describe the background solution with stabilized modulus (in a form similar to \cite{Kachru:2019dvo}) in section \ref{sec:background}, and then we consider fluctuations around that background in \ref{sec:fluctuations}, generalizing the correct ansatz for the unstabilized volume modulus.

Many calculational details are omitted here for convenience of presentation; we give them in appendix \ref{app:calculation}.

\subsection{Background}\label{sec:background}



Given that we wish to study the fluctuations of the stabilized volume modulus, it is important to mention the background of the supergravity solution that enables the stabilization. 
We can understand the background with gaugino condensate as a perturbation over a GKP background (geometry and flux). Let $\epsilon$ be the perturbation parameter, i.e., the strength of the gaugino condensate. As is typical (see for example \cite{Baumann:2008kq,Baumann:2010sx}), we assume that the equations of motion (and Bianchi identities) are those of the classical supergravity, at least away from local brane sources. 

We start with the metric ansatz 
\begin{equation}
ds^2=e^{2\Omega}e^{2A(y)}\hat{g}_{\mu\nu}(x)dx^{\mu}dx^{\nu}+ e^{-2A(y)}\tilde{g}_{mn}(y)dy^m dy^n. 
\label{eq:BackgroundMetric}
\end{equation}
where $\hat{g}_{\mu\nu}$ is the four-dimensional metric which, in the presence of the D7-brane gaugino condensate, is AdS as proposed in \cite{KKLT}. 
Further, $A(y)$ is the background warp factor including corrections due to the gaugino condensate, $\tilde{g}_{mn}$ is the internal space metric which need not be Ricci flat, and $e^{2\Omega}$ is the Weyl factor, defined as 
\begin{equation}\label{WeylDef}
    e^{-2\Omega}=\frac{\int d^6 y\sqrt{\tilde{g}}\,e^{-4A}}{\int d^6 y\sqrt{\tilde{g}}}. 
\end{equation}
The Weyl factor ensures that $\hat g_{\mu\nu}$ is the 4D Einstein frame metric once we allow moduli to fluctuate.
Note that $\hat g_{\mu\nu}$ depends on the noncompact $x$ coordinates even though it is a background; however, because it is maximally symmetric, neither the other fields or $\hat R$ will depend on $x$ in the background.

The D7-brane worldvolume theory contains couplings between the $G_{(3)}$ flux and the gaugino, so the gaugino condensate acts as a source for the flux \cite{Hamada:2018qef,Kallosh:2019oxv,Hamada:2019ack,Gautason:2019jwq}.
Therefore, the background flux $G_{(3)}$ also gets an imaginary anti-self dual part (and the ISD flux may also gain an $\O(\epsilon)$ modification):
\begin{align}
G_{(3)}&=G^+_{(3)} + G^-_{(3)} . 
\end{align}
Then, as pointed out in \cite{Baumann:2008kq,Baumann:2010sx}, we have to change the structure of $\tilde{F}_{(5)}$ such that it is not simply given by the corresponding warp factor but also involves a contribution $\alpha(y)$ that appears in the background 5-form as 
\begin{align}
    \Tilde{F}_{(5)} &= e^{4\Omega}\hat{\epsilon} \wedge \Tilde{d}e^{4A(y)}+e^{4\Omega}\hat{\epsilon} \wedge \Tilde{d}\alpha(y)+\Tilde{\star}\Tilde{d}e^{-4A(y)}-e^{-8A(y)}\tilde{\star}\Tilde{d}\alpha(y)\, .
\end{align}
One can also see using the equation of motion for the axiodilaton that if there is a presence of both ISD and IASD flux components, the axiodilaton is no longer a constant. Hence the background for the axiodilaton is now
\begin{equation}
    \tau(y) = \tau_0+\tau_1(y)+\cdots \equiv C_{(0)}(y)+ie^{-\phi(y)}
\end{equation}
where $\tau(y)$ satisfies the new background equation of motion order by order in the $\epsilon$. 
We can write the Einstein equations for this metric ansatz and the background form, fluxes and axiodilaton fields. The $(\mu,\nu)$ Einstein equation determines the background warp factor through the Poisson equation 
\begin{align}\label{eqn:warpfactorgauginobackground}
\tilde{\nabla}^{\tilde{2}}e^{-4A}=&-4e^{-8A}\partial_m A\partial^{\tilde{m}}\alpha-\frac{1}{2}e^{-12A}\partial_m \alpha \partial^{\tilde{m}}\alpha +e^{-4A}\tilde{R}+\frac{1}{2}e^{-2\Omega}e^{-8A}\hat{R}\nonumber\\
&-\frac{e^{\phi}}{2\times 3!}\left( G^{+}_{abc}\bar{G}^{+\widetilde{abc}} + G^{-}_{abc}\bar{G}^{-\widetilde{abc}}\right)-\frac{1}{2}e^{2\phi}e^{-4A}\partial_m \tau\partial^{\tilde{m}}\bar{\tau}+\text{local}.
\end{align}
We can see here that the warp factor no longer contains a modulus due to the presence of $A(y)$ on the RHS that cannot be canceled across the equation. 
(Since D7-brane gauge coupling depends on the warped volume of the 4-cycle the brane wraps and determines the value of the gaugino condensate, $G_{(3)}^\pm$ also depends directly on the warp factor and not just indirectly through backreaction in the equations of motion.)

The trace of the $(m,n)$ Einstein equation \eqref{eq:mnEinstein} gives the internal Ricci scalar $\tilde{R}$ 
\begin{equation}\label{eqn:internalcurvaturebackground}
    e^{4A}\tilde{R}=-\frac{3}{2}e^{-2\Omega}\hat{R}-\frac{1}{2}e^{-4A}\partial_j \alpha \partial^{\tilde{\jmath}}\alpha-4\partial_j A \partial^{\tilde{\jmath}}\alpha+\frac{1}{2}e^{4A}e^{2\phi}\partial_j \tau\partial^{\tilde{\jmath}}\bar{\tau} +\text{local}.
\end{equation}
Note that if we set $\alpha(y)$, $G^-_{(3)}$ and $\hat{R}$ to zero and $\tau(y)$ to a constant, the $(m,n)$ Einstein equation implies that $\tilde{R}$ is also zero, as expected. Under this substitution, the $(\mu,\nu)$ Einstein equation also reduces to the GKP equation for the warp factor.

Finally, the scalar component of the Bianchi identity for the five-form gives a Poisson equation for $\alpha(y)$, 
\begin{equation}\label{eqn:bianchibackground}
\tilde{\nabla}^{\tilde{2}}\alpha = e^{8A}\tilde{\nabla}^{\tilde{2}}e^{-4A} + 8 \tilde{\nabla}_a A \tilde{\nabla}^{\tilde{a}}\alpha + \frac{e^{8A+\phi}}{12}\left(G^{+}_{abc}\bar{G}^{+\widetilde{abc}} - G^{-}_{abc}\bar{G}^{-\widetilde{abc}}\right)+\text{local}. 
\end{equation}
We construct a simple linear combination of the equations (\ref{eqn:warpfactorgauginobackground}, \ref{eqn:internalcurvaturebackground}, \ref{eqn:bianchibackground}) to get the equation
\begin{equation}\label{eqn:background of master equation}
\tilde{\nabla}^{\tilde{2}}\alpha=-e^{-2\Omega}\hat{R}-e^{-4A}\partial_{m}\alpha \partial^{\tilde{m}}\alpha-\frac{e^\phi}{6}e^{8A}G_{abc}^- \bar{G}^{-\widetilde{abc}}+\textnormal{local}.
\end{equation}
This equation (with $\hat g_{\mu\nu}$ flat) first appeared in \cite{GKP}; when the local sources satisfy a BPS-like inequality, the final three terms on the right-hand side are negative semi-definite, which requires $\hat R$ to be zero or negative because the left-hand side integrates to zero.

Together with \eqref{eqn:warpfactorgauginobackground}, which we can re-write as
\begin{align}
\tilde{\nabla}^{\tilde 2} \left(e^{-4A}-\alpha\right) =&-8 e^{-8A}\partial_m A\partial^{\tilde m}\alpha -\frac{1}{12} e^{\phi}\left( G^+_{abc}\bar{G}^{+\widetilde{abc}}- G_{abc}^- \bar{G}^{-\widetilde{abc}}\right)+\textnormal{local} ,
\end{align}
and the equations of motion for $G_{(3)}$, \eqref{eqn:background of master equation} forms a nonlinear analog of an eigenvalue problem, which determines $\alpha$ and the warp factor as well as the external Ricci scalar $\hat R$, which is the analog of the eigenvalue.
Specifically, only discrete values of the external curvature will allow the warp factor and $\alpha(y)$ to remain smooth (or, if we excise the local sources as in \cite{Smith:2024ejf}, to have the proper boundary conditions at the excision). 
A simple explicit example occurs in warped Freund-Rubin-like compactifications with higher-dimensional flux and cosmological constant \cite{Kinoshita:2007uk}.
The description of the warp factor constraint and effective potential in \cite{Douglas:2009zn} is also similar.

As discussed in \cite{Baumann:2008kq,Baumann:2010sx}, there are two parametrically different cases for the backgrounds: $(\hat{R}, G^-, \alpha, \tilde{R})\sim\epsilon$ (case 1) vs $ (\hat{R}, \alpha, \tilde{R})\sim\epsilon^2$ and $G^-\sim \epsilon$ (case 2) ($y$ dependence of $\tau$ is $\sim\epsilon$ in both cases). 
In the first case, 
\begin{align}\label{eq:alphamaster1}
\tilde{\nabla}^{\tilde{2}}\alpha_1=-e^{-2\Omega_{0}}\hat{R}_1 +\textnormal{local}
\end{align}
to lowest order, so the integral relates $\hat R_1$ to local sources (or boundary conditions on $\alpha_1$ on an excision around the sources).
The lowest order in the second case is
\begin{align}\label{eq:alphamaster2}
\tilde{\nabla}^{\tilde{2}}\alpha_2=-e^{-2\Omega_{0}}\hat{R}_2 -\frac{e^{\phi_0}}{6}e^{8A_0}G_{1,abc}^- \bar{G}^{-\widetilde{abc}}_1+\textnormal{local},
\end{align}
so there is an additional contribution due to the integral of the IASD flux.


\subsection{Fluctuations}\label{sec:fluctuations}
In order to find the mass of the volume modulus, we need to consider fluctuations of the background, modifying the solution for the volume modulus that we reviewed in the introduction. Throughout, we work to first order in fluctuations around the stabilized background of section \ref{sec:background}.
In general, finding the mass eigenmodes of fluctuations that solve all the constraints (that is, all of the higher-dimensional equations of motion) requires knowing the background in detail, but it can nonetheless be possible to determine the mass by solving a subset of the equations of motion. 
For example, as we see in appendix \ref{app:frbreathing}, showing that the external components of the Einstein equation for a fluctuating volume modulus in a Freund-Rubin compactification vanish requires knowing the value of the stabilized modulus, but the internal Einstein equation determines the mass.
Similarly, solving all the Einstein equations for a stabilized radion mode in a Randall-Sundrum model requires knowledge of the background at the stabilized value of the modulus \cite{Csaki:2000zn}.
Since there is no detailed understanding of the full background with gaugino condensate yet, we do not attempt to solve all components of the equations of motion for the fluctuations. 
Instead, we will find a generalization of \eqref{eqn:background of master equation} that depends only on a few components and is robust against many variations of fluctuation ansatz.

We start with the fluctuation ansatz
\begin{align}
    e^{-4A(x,y)}=e^{-4A(y)} + \delta f(y)\, c(x) 
\end{align}
for the warp factor, where $c(x)$ is the volume modulus and $\delta f(y)$ is the extra-dimensional wave function of the volume modulus. 
Up to first order in fluctuations, we write the metric and the five-form flux as (see \cite{Cownden:2016hpf,Frey:2025rvf})
\begin{align}
ds^2&=e^{2\Omega(x)}e^{2A(x,y)}\hat{g}_{\mu\nu}(x)dx^{\mu}dx^{\nu}+2e^{2\Omega(x)}e^{2A(x,y)}\partial_\mu B_m(x,y)dx^\mu dy^m \label{eq:FluctMetric}\\
    &+ e^{-2A(x,y)}\tilde{g}_{mn}(y)dy^m dy^n ,\nonumber\\
    \tilde{F}_{(5)} &= \tilde
    {\star}\tilde{d}e^{-4A(x,y)}-e^{2\Omega(x)}\hat{d}(\tilde{\star}\tilde{d}L(x,y))+e^{4\Omega(x)}\hat{\epsilon}\wedge \tilde{d}\left(e^{4A(x,y)}+\alpha(x,y)\right)\nonumber\\
    &-e^{-8A(x,y)}\tilde{\star}\tilde{d}\alpha(x,y)+e^{4\Omega}\hat\star\hat dB\wedge\tilde d\left(e^{4A(x,y)}+\alpha(x,y)\right)
    -e^{2\Omega(x)}\hat{\star}\hat{d}\tilde{d}L(x,y) ,\label{eq:FluctF5}
\end{align}
where $L$ and $B$ are one-form compensators. We write $B_m$ as
\begin{equation}
    B(x,y)=c(x)\left(\tilde{d}\delta K(y)+ \delta B'(y)\right) 
\end{equation}
in terms of an exact part $\tilde{d}\delta K$ and a co-closed part $\delta B'$. 
Further, up to first order in fluctuations we have $\alpha(x,y)=\alpha(y)+\delta \alpha(y)\,c(x)$. 
We can also expand the Weyl factor $\Omega(x)=\Omega + \delta \Omega\, c(x)$, which is also conveniently written 
 \begin{equation}\label{WeylDef2}
 e^{-2\Omega(x)}=e^{-2\Omega}+\delta N c(x) ,\quad \delta N\equiv \frac{\int d^6 y \sqrt{\tilde{g}}\,\delta f(y)}{\int d^6y \sqrt{\tilde{g}}}.
\end{equation}
It is important to note that our ansatz does not allow $\tilde g_{mn}$ to fluctuate (so the denominators in (\ref{WeylDef},\ref{WeylDef2}) are constant).In the absence of the gaugino condensate the volume modulus is massless, and its wavefunction is constant on the internal space $\delta f(y) = 1$, $\delta \alpha(y)=0$, as reviewed in Section \ref{sec:WarpedVolModulus}.

Before considering the equations of motion, we pause to comment on the ansatz (\ref{eq:FluctMetric},\ref{eq:FluctF5}).
Like the classical case for the (unstabilized) volume modulus (i.e., when there is no gaugino condensate), we use a compensator in the $(\mu,m)$ metric component to satisfy the constraints, which also appears in $\tilde F_{(5)}$.
The compensator terms are somewhat more complicated than those for the unstabilized volume modulus. First, as noted, $B_m$ is no longer necessarily exact.
Furthermore, compensators appear in several components of $\tilde F_{(5)}$. Both of these points are related to the fact that we expect the fluctuation of $e^{-4A}$ to depend on both $x$ and $y$ coordinates, so $\hat d \tilde\star \tilde de^{-4A}\neq 0$. 
Then the $\hat{d}(\tilde{\star}\tilde{d}L)$ term satisfies the component of the Bianchi identity with one $x$ index and five $y$ indices (this constraint is a Poisson equation for $L$; see \eqref{eq:Lpoisson}), and the remaining component containing $L$ is required for self-duality.
The remaining terms involving $B$ are also required for self-duality since the background components of the 5-form are not self-dual since the metric is no longer block diagonal.
In combination with the Bianchi identity, the $(\mu,m)$ component of the Einstein equation determines $B$ as in \eqref{eq:BminusL}.
We do not attempt to solve the component of the Bianchi identity with 4 legs along $x$ and 2 legs on $y$. However, none of these equations contribute to the master equation \eqref{eqn:mastereqn} below.
We have checked that an ansatz which solves the Bianchi identity with 2 legs on $y$ rather than the component with 5 legs on $y$ gives the same equation \eqref{eqn:mastereqn} up to changes in the order unity numerical coefficients, so we consider our results robust up to order one numerical factors.

We must also consider whether $\hat g_{\mu\nu}$ remains maximally symmetric or acquires other fluctuations; the external Ricci scalar $\hat R$ is particularly important due to its appearance in many of the equations of motion.
We have seen that we can consider $\hat R$ as a nonlinear analog to the smallest eigenvalue of some system. Similar to an eigenvalue, it stands to reason that it may be the minimum value of some functional of the supergravity background. In that case, any change to $\hat R$ will be quadratic in any perturbation of the background.
This logic is consistent with an interpretation in effective field theory; a perturbation of the background reduces to a scalar field, and the stress tensor (above the cosmological constant) is quadratic in the scalar.

For the ansatz \eqref{eq:FluctMetric}, the $(\mu,\nu)$ Einstein equation gives a constraint 
\begin{equation}\label{eq:Klaplace}
    -2\hat{\nabla}_\mu \hat{\nabla}_\nu \Omega(x) + 4\hat{\nabla}_\mu \hat{\nabla}_\nu A(x,y) +e^{4A(y)}e^{2\Omega}\hat{\nabla}_\mu \hat{\nabla}_\nu\tilde{\nabla}^{\tilde{a}}B_{a}(x,y)=0 .
\end{equation}
Up to first order in fluctuations, this is 
\begin{equation}\label{eqn:firstflucmunuoffdiagonal}
   \delta N e^{2\Omega}-\delta fe^{4A}+e^{4A(y)}e^{2\Omega}\tilde{\nabla}^{\tilde{2}}\delta K(y)=0.
\end{equation}
The generalization of \eqref{eqn:background of master equation} valid through first order in the fluctuations is
\begin{align}\label{eqn:mastereqn}
\tilde{\nabla}^{\tilde{2}}\alpha(x,y)=&-e^{-2\Omega(x)}\hat{R}-e^{-4A(x,y)}\partial_j \alpha(x,y) \partial^{\tilde{\jmath}}\alpha(x,y)-\frac{1}{6}e^{\phi(x,y)} e^{8A(x,y)}G_{abc}^-(x,y) \bar{G}^{-\widetilde{abc}}(x,y)\nonumber\\
&+3e^{-2\Omega}\left(3\hat{\nabla}^{\hat{2}}\Omega(x,y)-2\hat{\nabla}^{\hat{2}}A(x,y)\right)\nonumber\\
&-\frac{5}{2}e^{4A(y)}\hat{\nabla}^{\hat{2}}\tilde{\nabla}^{\tilde{a}}B_{a}(x,y)-4e^{4A(y)}\tilde{\nabla}^{\tilde{a}}A(y) \hat{\nabla}^{\hat{2}}B_{a}(x,y)+\textnormal{local}\nonumber\\
=&-e^{-2\Omega(x)}\hat{R}-e^{-4A(x,y)}\partial_j \alpha(x,y) \partial^{\tilde{\jmath}}\alpha(x,y)-\frac{1}{6}e^{\phi(x,y)} e^{8A(x,y)}G_{abc}^-(x,y) \bar{G}^{-\widetilde{abc}}(x,y)\nonumber\\
&+4e^{-2\Omega}\hat\nabla^{\hat 2}\left(\Omega(x)+A(x,y) \right)-4e^{4A(y)}\tilde{\nabla}^{\tilde{a}}A(y) \hat{\nabla}^{\hat{2}}B_{a}(x,y)+\textnormal{local}
\end{align}
after using \eqref{eq:Klaplace}.
In comparison, \cite{Kachru:2019dvo} derived a similar equation (their (2.26)) for the volume modulus using the scaling ansatz \eqref{eq:unwarpedVolModulus}, which as we have seen does not correspond to the volume modulus in warped spaces. A similar expression can also be found in (5.29) of \cite{Giddings_Maharana}. We see two important differences due to the proper identification of the volume modulus as a shift of the warp factor and the required inclusion of compensators: the various terms on the right-hand side scale differently with the volume modulus (contributing to a potential that now stabilizes the volume modulus), and terms with the d'Alembertian of the modulus now depend on the internal position $y$. The latter effect in particular impacts our determination of the mass, and could in principle lead to a dependence of the mass on warping when solving the resulting eigenvalue problem.

To find a mode of definite 4D mass, we replace $\hat\nabla^{\hat 2}\to m^2$ in \eqref{eqn:mastereqn}, 
which simplifies considerably if we expand in powers of the gaugino condensate.
While we have presented the spacetime-dependent modulus as an expansion in fluctuations around the background of the previous subsection, this expansion commutes with the expansion in the gaugino condensate.
As a result, it is also useful to think of the above as an expansion of the solution for the spacetime-dependent modulus (as reviewed in the introduction) in $\epsilon$, and we will take this point of view while determining the mass.


We begin by examining case 1 for the parametric dependence ($(\hat{R}, G^-, \alpha, \tilde{R})\sim\epsilon$), noting that $m^2\sim\epsilon$ as well. 
Without the gaugino condensate, the wavefunction of the volume modulus is $\delta f_0(y)=1$, so $\delta A_0(y) =-e^{4A_0(y)}/4$, as we recall from the introduction (see \cite{FTUD} also for the following).
Further, this implies $\delta N= 1 + \mathcal{O}(\epsilon)$, so $\delta\Omega_0=-e^{2\Omega_0}/2$. 
The co-exact one form $\delta B'_0=0$ for the volume modulus without gaugino condensate \cite{FTUD}, so it does not contribute at lowest order.
In principle, there can be an $\mathcal{O}(\epsilon)$ contribution to the compensator $\delta K$, but it is also not relevant since it will only appear in \eqref{eqn:mastereqn} at higher order in $\epsilon$.
In the end, we find \eqref{eqn:mastereqn} at leading order in $\epsilon$ becomes
\begin{equation}\label{eqn:mastereqnfirstorder}
\tilde{\nabla}^{\tilde{2}}\delta\alpha_1 = -\hat{R}_1-2(m^2)_1-e^{-2\Omega_0}e^{4A_0(y)}(m^2)_1 - (m^2)_1 \tilde{\nabla}_j \delta K_0(y) \tilde{\nabla}^{\tilde{\jmath}}\left(e^{4A_0(y)}\right)
+\delta\textnormal{local} .
\end{equation}
As explained above, we assume that there is no variation in $\hat R$ to first order in the fluctuations.
The key property of \eqref{eqn:mastereqnfirstorder} is that it decouples from the fluctuations of other supergravity fields except for the compensator $\delta K$ evaluated for vanishing gaugino condensate, which can in principle be calculated for a classical compactification with ISD flux. 
That makes \eqref{eqn:mastereqnfirstorder} a powerful test of the volume modulus.

In case 2 ($ (\hat{R}, \alpha, \tilde{R})\sim\epsilon^2$ and $G^- \sim \epsilon$), the first order of \eqref{eqn:mastereqn} in $\epsilon$ is the same as \eqref{eqn:mastereqnfirstorder} with $\delta\alpha_1=0$ and $\hat R_1=0$, which implies that $m^2$ vanishes at first order. At second order in $\epsilon$,
\begin{align}\label{eqn:mastereqncase2}
\tilde{\nabla}^{\tilde{2}}\delta\alpha_2 =& -\hat{R}_2-2(m^2)_2-e^{-2\Omega_0}e^{4A_0(y)}(m^2)_2 -(m^2)_2 \tilde{\nabla}_j \delta K_0(y) \tilde{\nabla}^{\tilde{\jmath}}\left(e^{4A_0(y)}\right)
\nonumber\\
&+\frac 13 e^{\phi_0}e^{8A_0(y)}\left(e^{4A_0(y)}G^-_{abc}\bar G^{-,\widetilde{abc}}-\frac 12 G^-_{abc}\delta\bar G^{-,\widetilde{abc}}-\frac 12 \delta G^-_{abc}\bar G^{-,\widetilde{abc}}\right) +\delta\textnormal{local},
\end{align}
where we have recalled that $\tau$ and $\tilde g_{mn}$ do not fluctuate at $\O(\epsilon^0)$, and the first term in the second line is due to the fluctuation of the warp factor. 
In this case, the mass depends on the 3-form fluctuations, so the various supergravity fields remain coupled, and it is not possible to analyze this equation independently in general.


As a reminder, we will use only the master equation for $\alpha(x,y)$ to estimate the mass of the volume modulus (at the lowest nontrivial order in the gaugino condensate, as represented in (\ref{eqn:mastereqnfirstorder},\ref{eqn:mastereqncase2})). 
This equation follows from the external Einstein equation, the trace of the internal Einstein equation, and the scalar component of the 5-form Bianchi identity, but we do not attempt to solve these equations independently with the gaugino condensate. 
We also do not attempt to solve the trace-free part of the internal Einstein equation or the component of the Bianchi identity with four external legs; adjusting the 5-form ansatz to solve the latter equation leaves the form of \eqref{eqn:mastereqn} unchanged but alters some of the numerical coefficients (while leaving them of order unity).

\section{Simplified Models}\label{sec:Simplified Toy Models}
In the previous section we analyzed the 10-dimensional Einstein equations for fluctuations of the warped volume modulus about the backgrounds of \cite{KKLT}. After imposing constraints, the mass of the warped volume modulus is determined by solutions of the eigenvalue-like equation (\ref{eqn:mastereqnfirstorder}) or (\ref{eqn:mastereqncase2}). 

In order to solve this equation for the mass we need to specify the background geometry, including the warp factor. 
In this section, we will consider solving (\ref{eqn:mastereqnfirstorder}, \ref{eqn:mastereqncase2}) in several different backgrounds that will allow us to test the robustness of our conclusions.
First, we consider a weakly warped, large-volume limit. 
We will then consider our background to consist of an $\text{AdS}_5\times X_5$ warped throat, as a simple model of warped regions in realistic compactifications \cite{KSThroat}, similar to the Randall-Sundrum model \cite{Randall:1999ee}.
Finally, we will consider an $\mbox{AdS}_5\times X_5$ throat glued onto a flat bulk spacetime, which will allow us to interpolate between the strongly warped and large volume (weakly warped) limits.


\subsection{Large Volume Limit} \label{sec:largeVolume}

Let us consider the large volume/weakly warped limit in which
the volume modulus is stabilized with $e^{-4A_0(y)}$ approximately constant and large, so that $e^{-2\Omega_0} \approx e^{-4A_0}\gg 1$ from (\ref{WeylDef}).
The background equations (\ref{eq:alphamaster1},\ref{eq:alphamaster2}) both take the same form
\begin{equation}
    \tilde \nabla^2 \alpha = -e^{-2\Omega_0} \hat R + \textnormal{local}\, ,
    \label{eq:masteralphaLargeVolLimit}
\end{equation}
since contributions to (\ref{eq:alphamaster2}) from $G^-$ flux are suppressed in this limit as the large volume dilutes the fluxes away.
We will drop the numerical subscript, since the equations for both parametric cases take the same form.
For simplicity, we will take the internal space to be flat space parameterized by a linear ``radial" interval $0 \leq r \leq r_*$, with the other directions described by compact angular coordinates at fixed size.
The local terms here are sourced by the gaugino condensation, as in \cite{Kachru:2019dvo}, and given in terms of $\hat R$ through the integral of (\ref{eq:masteralphaLargeVolLimit}). We will capture the effects of the local sources as a Neumann boundary condition located at $r_*$, which we denote as $a = \partial_r \alpha(r_*)$ (with vanishing Neumann boundary conditions at $r=0$, $\partial_r \alpha(0) = 0$).
Integrating (\ref{eq:masteralphaLargeVolLimit}) gives $a = -r_* e^{-2\Omega_0}\hat R$, evaluated on the background.

For fluctuations, the compensator $K_0$ determined by (\ref{KCompensator}) becomes $K_0\approx \mbox{const}$, so that terms proportional to $\tilde \nabla K_0$ and $\tilde \nabla e^{4A_0}$ become vanishingly small.
Together with the suppression of terms from flux, the eigenvalue equations (\ref{eqn:mastereqnfirstorder}, \ref{eqn:mastereqncase2}) also take the same form in the large volume/weakly warped limit
\begin{equation}
    \tilde \nabla^2 \delta \alpha = - \hat R - 3 m^2 + \delta(\textnormal{local}) \, .
    \label{eqn:mastereqnLargeVolLimit}
\end{equation}
Integrating (\ref{eqn:mastereqnLargeVolLimit}) then gives the mass eigenvalue in the large volume limit
\begin{equation}
m^2 = -\frac{1}{3} \left(\hat R + \frac{\delta a}{r_*}\right)\,,
\label{eq:MassLargeVolume}
\end{equation}
in terms of the 4-dimensional curvature and the boundary condition $\delta a = \partial_r \delta \alpha(r_*)$.
In order to calculate $\delta a$, we would need to know the full moduli dependence of the gaugino condensation local terms off-shell.
However, we can estimate the contribution of $\delta a$ to the mass (\ref{eq:MassLargeVolume}) by taking the variation of the boundary condition $a$ given above. Since the $e^{-2\Omega_0}$ factor contains the dominant volume modulus dependence, we find $\delta a \sim {\mathcal O}(1)\, r_*\, \hat R$. 
Altogether, then, we expect  $m^2 \sim -{\mathcal O}(1)\, \hat R$ in the large volume limit. Interestingly, there does not appear to be any volume suppression or other parametric dependence that can give rise to a scale separation between the volume modulus mass and the 4-dimensional curvature scale.
We will return to this issue in Section \ref{sec:discuss}.

\subsection{Warped throat model}\label{sec:throat}
The first background geometry is that of an AdS throat, similar to the Randall-Sundrum 2-brane model \cite{Randall:1999ee}. The metric is given by 
\begin{equation}\label{eqn:adsthroatmetric}
    ds^2=e^{2\Omega_0}e^{2A_0(r)}\hat{g}_{\mu\nu}dx^{\mu}dx^{\nu}+e^{-2A_0(r)}(dr^2 + r^2 d\Omega^2_{T^{1,1}})
\end{equation}
where $e^{-4A_0}\equiv\ell^4/r^4$. The $r$ direction runs from $r_0$ to $r_c$ (note that the $0$ subscript on $r_0$ is not indicative of $\epsilon$ dependence) with $r_c$ defined so $e^{-4A_0(r_c)}=\ell^4/r_c^4=c$, where $c$ is the expectation value of the volume modulus.
As a result, $r_c$ fluctuates with the volume modulus (and can be taken as an alternate definition of the modulus in this model).
The other five spatial directions are compact and are in the form of a $T^{1,1}$ manifold. 
This background is a common approximation for the Klebanov-Strassler throat \cite{KSThroat}, and we express the gaugino background as a perturbation of this classical background.
We collect a few common formulas here and then discuss the two cases for scaling with the gaugino condensate separately.

We can calculate 
\begin{equation}\label{eq:weyl1}
e^{-2\Omega_0}=\frac{\Omega_5\int_{r_0}^{r_c} \frac{\ell^4}{r^4}r^5 dr}{\Omega_5\int_{r_0}^{r_c} r^5 dr}
=\frac{\ell^4(r_c^2-r_0^2)/2}{(r_c^6-r_0^6)/6}=\frac{3\ell^4}{r_0^4+r_0^2 r_c^2 +r_c^4}. 
\end{equation}
In the long throat limit $r_0\to 0$, this becomes $e^{-2\Omega_0}\approx 3\ell^4/r_c^4$. 
Then the fluctuation $\delta r_c = -r_c^5/4\ell^4\approx -r_c e^{2\Omega_0}$.

We can also find the compensator $\delta K_0$.
The equation (\ref{eqn:firstflucmunuoffdiagonal}), at zeroth order in the gaugino, can be written as
\begin{equation}
    \tilde{\nabla}^{\tilde{2}}(\delta K_0)=\frac{d^2 (\delta K_0)}{d r^2}+\frac{5}{r}\frac{d(\delta K_0)}{dr} =e^{-2\Omega_0}-e^{-4A_0(y)}.
\end{equation}
We assume Neumann boundary condition at $r_0$, which gives a smooth function in the limit $r_0\to 0$ when there is rotational symmetry on $T^{1,1}$.
We find 
\begin{equation}\label{eqn:compensator in throat}
\delta K_0(r)=\frac{\ell^4}{4r^2}+\frac{1}{12}e^{-2\Omega_0}r^2 -\frac{1}{4r^4}\left(\frac{\ell^4 r_0^2}{2}-\frac{1}{6}e^{-2\Omega_0}r_0^6\right)+c_{\text{comp}}
\end{equation}
where $c_{\text{comp}}$ is fixed by taking Robin boundary conditions at $r_c$. Because we need only the derivative of $\delta K_0$, the details
of the boundary conditions at $r_c$ are unimportant.

We will not include local sources but model their physics through boundary conditions at the UV end of the throat $r=r_c$.

\subsubsection{Parametric scaling case 1}
Considering the parametric dependence as $((\hat{R},G^-,\alpha,\tilde{R})\sim \epsilon)$, as presented in the section (\ref{sec:fluctuations}), we start by estimating the Ricci scalar using \eqref{eq:alphamaster1} for the 5-form function $\alpha$.
As for the compensator $\delta K_0$, we assume that $\alpha$ satisfies standard Neumann boundary conditions at $r_0$. 
We assume that there are no local sources in the throat, but we model their appearance in the compactification by choosing nontrivial Neumann boundary conditions $\alpha'(r_c)=a$.\footnote{To include local sources at a position $r_b$, we would also include a jump condition on $\alpha'$ at $r_b$ similar to a delta function potential in the Schr\"odinger equation.} 
It is important to understand these boundary conditions. 
Because $\alpha(x,y)$ only appears when acted on by $\tilde d$, a constant background $\alpha$ or $\delta\alpha$ is pure gauge. Therefore, choosing Dirichlet or mixed (Robin) boundary conditions is physically meaningless.
Solving \eqref{eq:alphamaster1}, we see that 
\begin{equation}\label{eq:alpharicci1}
a_1 = -\frac{1}{6} \left(r_c - \frac{r_0^6}{r_c^5} \right) e^{-2\Omega_0}\hat R_1\, ,
\end{equation}
which relates the Ricci scalar to the boundary conditions, precisely as expected from our discussion in section \ref{sec:background}.
It is important to note that \eqref{eq:alpharicci1} applies on the background, but $a_1$ is presumably a function of the gaugino condensate and not the bulk fields.
Therefore, the variations of the two sides of \eqref{eq:alpharicci1} are different, but the natural expectation is that they are similar in size.

In this case, the mass squared is also of order $\epsilon$. Then the fluctuations of $\alpha(x,y)$ at lowest order in $\epsilon$ are governed by equation (\ref{eqn:mastereqnfirstorder}), which is 
\begin{equation}
    \frac{d^2 (\delta \alpha_1)}{dr^2}+\frac{5}{r}\frac{d(\delta\alpha_1)}{dr} = -\hat{R}_1-2(m^2)_1-e^{-2\Omega_0}e^{4A_0(y)}(m^2)_1 - 4e^{4A_0(y)}(m^2)_1 \partial_r \delta K_0 \partial_r A_0
\end{equation}
in the warped throat model.
We must also set the boundary conditions; similar to the background, we take Neumann conditions at both boundaries with a nonvanishing derivative at $r_c$:
\begin{equation}\label{eqn:boundcond in pure throat}
\delta \alpha_1'(r_0)=0 , \quad \delta \alpha_1'(r_c)=\delta a_1.
\end{equation}
This boundary condition should be interpreted to include the effects of the fluctuation of the boundary of the compact space itself (since $r_c$ changes with the modulus).
To estimate the value $\delta a_1$, we consider the fluctuation of the right-hand side of \eqref{eq:alpharicci1}, which is
\begin{equation}\label{eq:bcestimate1}
-\frac 16\hat R_1\left[ \left( r_c -\frac{r_0^6}{r_c^5}\right) +e^{-2\Omega_0} \delta r_c \left(1+ \frac{5r_0^6}{r_c^6}\right)\right] .
\end{equation}
As a rough estimate, we will assume $r_0\ll r_c$ and ignore factors of order unity (including allowing a change of sign), so we will use
\begin{equation}
\delta a_1\approx \pm e^{2\Omega_0} a_1 \approx \pm r_c \hat R_1 \, .
\end{equation}
In the other models we consider below, we will also see that $\delta a\approx \pm e^{2\Omega_0}a$ is a reasonable estimate, which makes $\delta a$ the 4D Ricci scalar times a natural length scale of the compactification.

Using the derived function for $\delta K_0(r)$ in equation (\ref{eqn:compensator in throat}), the mass eigenvalue is 
\begin{equation}\label{eqn:purethroatcomplicatedmass}
     (m^2)_1=\frac{\ell^4 e^{2\Omega_0} \left[-\hat{R}_1 \left(r_c^6-r_0^6\right)-6 \delta a_1 r_c^5\right]}{3 \ell^4 e^{2\Omega_0} r_0^2 \left(r_c^4-r_0^4\right)+r_c^4 \left(r_c^6-r_0^6\right)} .
 \end{equation}
On substituting the value for $e^{-2\Omega_0}$ as derived in equation (\ref{eq:weyl1}), we find 
\begin{equation}
     (m^2)_1=-\frac{\hat{R}_1}{3}-\frac{2 \delta a_1 r_c^5}{ \left(r_c^6-r_0^6\right)} . 
 \end{equation}
It is important to note that the mass goes to a constant given by the spacetime curvature in the long throat limit rather than scaling with the warp factor.  
Assuming that the mass is not tachyonic implies a constraint $\delta a_1 \leq -r_c\hat R_1(1-r_0^6/r_c^6)/6$.

\subsubsection{Parametric scaling case 2}
Next we consider the second parametric dependence ($ (\hat{R}, \alpha, \tilde{R})\sim\epsilon^2$ while $G^- \sim \epsilon$). 
In this case, we need to include effects of the AISD flux in the throat.

To model the $G^-_1$ in the AdS throat, we follow \cite{Baumann:2010sx}, which gives  
\begin{equation}
    G^-_1=-2i\frac{\delta-2}{\delta}r^{\delta-4}\left(\tilde d \Omega_2+\delta\frac{\tilde d r}{r}\wedge\Omega_2 \right)
\label{eq:GminusFlux}
\end{equation}
where $\Omega_2$ is a two form on the angular directions. The constant $\delta (\neq0)$ denotes a choice of mode in the AdS throat spectrum. 
Depending on the chosen components of $G^-_1$, the allowed values of $\delta$ are tabulated in \cite{Baumann:2010sx}. 
With this flux, 
\begin{equation}
e^{\phi_0} G^-_{1,abc} \bar G_1^{-\widetilde{abc}} \equiv 3 g_1^2 (\delta -2)^2 r^{2\delta -14},
\end{equation}
where $g_1$ is a factor that includes $e^{\phi_0}$, numerical factors coming from $G^-_1$, and the square of the $\Omega_2$ form.

We also need to determine the boundary conditions on $\alpha(x,y)$. As in the other scaling case, we have
\begin{equation}\label{eq:a2throat}
\alpha'_2(r_c) = a_2 = -\frac 16 \left( r_c - \frac{r_0^6}{r_c^5} \right) e^{-2\Omega_0} \hat R_2 -\frac{g_1^2}{4} \frac{(\delta-2)^2}{2\delta} \frac{r_c^{2\delta-5}-r_0^{2\delta}/r_c^5}{\ell^8} .
\end{equation}
Again, we see how the 4D curvature relates to the flux and the boundary conditions (i.e., local terms) on the background solution. 
Using similar reasoning to the other scaling case, we take boundary conditions
\begin{equation}
\delta\alpha_2'(r_0)=0,\quad \delta\alpha_2'(r_c) =\delta a_2 \approx \pm e^{2\Omega_0}a_2 \label{eq:case2bc}
\end{equation}
for the fluctuations. We have again estimated $\delta a_2$ to be similar in size to the variation of the right-hand side of \eqref{eq:a2throat}.

Here, the mass squared is order $\epsilon^2$ and is determined by (\ref{eqn:mastereqncase2}). In the model of a single warped throat, it is
\begin{align}
     \frac{\mathrm d^2 (\delta \alpha_2)}{\mathrm d r^2}+\frac{5}{r}\frac{\mathrm (d\delta\alpha_2)}{\mathrm d r} =& -\hat{R}_2-2(m^2)_2-e^{-2\Omega_0}e^{4A_0(y)}(m^2)_2 - 4(m^2)_2 \partial_r \delta K_0(y) \partial_r\left(e^{4A_0(y)}\right)\nonumber\\
    &+\frac 13 e^{\phi_0}e^{8A_0(y)}\left(e^{4A_0(y)}G^-_{1,abc}\bar G^{-\widetilde{abc}}_1-\frac 12 G^-_{1,abc}\delta\bar G^{-,\widetilde{abc}}_1-\frac 12 \delta G^-_{1,abc}\bar G^{-,\widetilde{abc}}_1\right).\label{eq:case2throat1}
\end{align}
or
\begin{align}
    \frac{\mathrm d^2 (\delta \alpha_2)}{\mathrm d r^2}+\frac{5}{r}\frac{\mathrm d(\delta\alpha_2)}{\mathrm d r} =& -\hat{R}_2-2(m^2)_2-e^{-2\Omega_0}e^{4A_0(y)}(m^2)_2 - 4(m^2)_2 \partial_r \delta K_0(y) \partial_r\left(e^{4A_0(y)}\right)\nonumber\\
&+g_1^2(\delta-2)^2 \frac{r^{2\delta-2}}{\ell^{12}}.\label{eq:case2throat2}
\end{align}
We have absorbed the terms of \eqref{eq:case2throat1} that contain 3-form fluctuations $\delta G^-_{1}$ into the $g_1^2$ terms of \eqref{eq:case2throat2}, which implicitly makes the assumption that these terms have the same radial profile in the throat (see appendix \ref{app:form} for a rough argument in support of this assumption). 
If they do not, \eqref{eq:case2throat2} will have an additional term with a different power law in $r$, but our qualitative results for the mass will not change. 

Again taking $\delta K_0(r)$ from (\ref{eqn:compensator in throat}), we estimate the mass of the volume modulus using boundary conditions as in \eqref{eq:case2bc}. The mass eigenvalue is
\begin{equation}
    (m^2)_2=-\frac{\hat R_2}{3}-\frac{2 \delta a_2 r_c^5}{r_c^6-r_0^6}+\frac{g_1^2 (\delta-2)^2(r_c^{2\delta+4}-r_0^{2\delta+4})}{\ell^{12} (\delta+2)(r_c^6-r_0^6)}.
\end{equation}
From \eqref{eq:case2bc}, the $\delta a_2$ term is similar in magnitude to the combination of the first two terms.
In the long throat limit $r_0\to 0$, the mass eigenvalue reduces to 
\begin{equation}\label{eq:mass2throat}
     (m^2)_2\approx -\frac{\hat R_2}{3}-\frac{2\delta a_2}{r_c}+\frac{g_1^2 (\delta-2)^2 r_c^{2\delta-2}}{\ell^{12} (\delta+2)}.
\end{equation}
As in the other case, the mass goes to a constant even when warping becomes strong, and we also expect the three terms in \eqref{eq:mass2throat} to be similar in size.


\subsection{Throat and Bulk Model}\label{sec:throatandbulk}
As a more realistic model, we consider a model including both a throat and bulk with the extra dimensions factorizing into a radial direction and five angular directions.
The unwarped geometry is a conifold (radial direction with angular directions forming a $T^{1,1}$ manifold) truncated at fixed values $r_0$ and $r_m$ in the radial direction.
This is the same metric as \eqref{eqn:adsthroatmetric} with $r_0\leq r\leq r_m$.
We take the warp factor to be $e^{-4A_0} = c+\ell^4/r^4$ with $c$ a constant and further simplify by using a piecewise approximation $e^{-4A_0} =\ell^4/r^4$ for $r\leq r_c$ and $e^{-4A_0} =c$ for $r\geq r_c$ with $r_c$ defined so $\ell^4/r_c^4\equiv c$. In string length units, $r_0$ is small, whereas $r_m$ is order one to ten.
For this geometry, the Weyl factor is
\begin{equation}\label{eqn:OmegaBulkThroat}
    e^{-2\Omega_0}=\frac{\Omega_5\int_{r_0}^{r_c}\frac{\ell^4}{r^4}r^5 dr+\Omega_5\int_{r_c}^{r_m} \frac{\ell^4}{r_c^4}r^5 dr}{\Omega_5\int_{r_0}^{r_m}r^5 dr}=
    \ell^4\frac{r_m^6/r_c^4+2r_c^2-3r_0^2}{r_m^6-r_0^6} .
\end{equation} 
When $r_0\ll r_c$, the throat region is long, so we take the long throat limit as $r_0\to 0$. For a fixed $r_0$, the maximum warped volume (and Weyl factor) is $e^{-2\Omega_0}=\ell^4/r_0^4$, which occurs for $r_c\to r_0$ and diverges as $r_0\to 0$. 
To recap the large volume regime: for a sufficiently large but finite (stabilized) value $c$ for the volume modulus, $r_c=\ell c^{-1/4} \leq$ the naive value of $r_0$. Then we have instead $r_0=r_c$. In the infinite volume limit $c\to\infty$, we also have $r_0\to 0$.
We also note that fluctuation of the volume modulus gives
\begin{equation}
\delta r_c = -\frac{r_c^5}{4\ell^4} \approx -r_c e^{2\Omega_0}
\end{equation}
for either $r_m\gg r_c$ or the large volume regime $r_c\approx r_0$, so we take this approximation to hold up to factors of order unity for interesting regions of parameter space.

The $G^-_1$ flux in the throat is given by \eqref{eq:GminusFlux}, as in our model with only a throat. 
In the bulk, it is frozen to the value it has on the throat-bulk intersection, i.e., $G^-_1(r)=G^-_1(r_c)$.
We also limit local sources to the junction between throat and bulk ($r=r_c$) and the end of the bulk ($r=r_m$). 
Effectively, this means that the radial direction we consider is actually a path through the extra dimensions, and the D7-branes that host the gaugino condensate are separated from that path along the angular directions except \af{at $r_m$}.

We also must solve the compensator equation (\ref{eqn:firstflucmunuoffdiagonal}) in the throat and the bulk. 
In the throat, the compensator is given by (\ref{eqn:compensator in throat}) with Neumann boundary conditions for $\delta K_0$ at $r_0$ and $c_{\text{comp}}$ determined by the matching conditions at $r_c$. 
However, we note that in equations (\ref{eqn:mastereqnfirstorder}) and (\ref{eqn:mastereqncase2}), all the terms that multiply the compensator $\delta K_0(y)$ are also multiplied by the derivative of the warp factor. As the warp factor in the bulk is a constant, those terms in the equation for the calculation of $\delta\alpha$ for the bulk are absent. As a result, we do not need to determine $\delta K_0$ in the bulk.

As in the model of the previous section, we will model local sources through boundary conditions at $r_m$. 
Local sources at points $r=r_b$ (in either the throat or bulk) introduce a specified discontinuity in $\alpha'(r)$, much like a delta function potential in the Schr\"odinger equation. 
This introduces algebraic complication without new physics or new qualitative features to the mass eigenvalue, so we ignore this possibility here. If the D7-branes hosting the gaugino condensate extend into the warped throat (for example, if we take local sources at
$r_b< r_c$), the value $\epsilon$ of the gaugino condensate will depend on the warp factor. However, the condensate is still fixed in the
background and related to other parameters of the compactification (such as $\hat R$) through the equations of motion as we have
described. Of course, fluctuations in the volume modulus and warp factor will cause the condensate to fluctuate, which we
model through boundary conditions on $\delta\alpha$ below. Our main interest is other effects of the warp factor on the modulus:
does the background warp factor away from the branes affect the profile of the modulus?

\subsubsection{Parametric scaling case 1}
First consider the parametric dependence $(\hat{R},G^-,\alpha,\tilde{R})\sim \epsilon$. 
Using the same logic as for the warped throat model to replace local sources with boundary conditions, we see that the boundary condition on the background $\alpha_1(r)$ at $r=r_m$ is
\begin{equation}
a_1 = -\frac 16 \left(r_m-\frac{r_0^6}{r_m^5}\right)e^{-2\Omega_0}\hat R_1 .
\end{equation}
We again assume the boundary condition $\delta\alpha_1(r_m)=\delta a_1$ is of the same order as the variation of the right-hand side, which yields
\begin{equation}
\delta a_1 \approx \pm r_m\hat R_1 \approx \pm a_1 e^{2\Omega_0} 
\end{equation}
which is roughly a string length times $\hat R_1$.
In principle, we could include discontinuities in $\alpha'(x,r)$ to model branes with gaugino condensates at internal points $r_0<r<r_m$; we have checked that this changes the quantitative results in detail but not qualitative or order of magnitude results.

The equation determining $\delta \alpha_1$ is
\begin{equation}
    \frac{\mathrm d^2 (\delta \alpha_1)}{\mathrm d r^2}+\frac{5}{r}\frac{\mathrm d(\delta\alpha_1)}{\mathrm d r} = -\hat{R}_1-2(m^2)_1-e^{-2\Omega_0}e^{4A_0(y)}(m^2)_1 - 4e^{4A_0(y)}(m^2)_1 \partial_r \delta K_0 \partial_r A_0
\end{equation}
in the throat and
\begin{equation}
    \frac{\mathrm d^2 (\delta \alpha_1)}{\mathrm d r^2}+\frac{5}{r}\frac{\mathrm d(\delta\alpha_1)}{\mathrm d r}= -\hat{R}_1-2(m^2)_1-e^{-2\Omega_0}e^{4A_0(y)}(m^2)_1
\end{equation}
in the bulk. Solving,
\begin{align}
   \delta\alpha_{1,\text{throat}}&=-\frac{t_{\text{I}}\ell^4}{4 r^4}+t_{\text{II}}-\frac{r^2 \hat{R}_1}{12\ell^2}-r^6(m^2)_1\frac{ (r_m^6+2r_c^6 -3r_0^2 r_c^4)}{36\ell^6 r_c^4 (r_m^6-r_0^6)}\nonumber\\
   &\hphantom{=,} +\log{(r/\ell)}(m^2)_1\frac{ r_0^2 (2r_0^4 r_c^6 + r_0^4 r_m^6 -3 r_c^4r_m^6)}{6 r_c^4(r_m^6-r_0^6)}\\
   \delta\alpha_{1,\text{bulk}}&=b_{\text{I}} -\frac{b_{\text{II}}\ell^4}{4r^4}  + \frac{r^2}{12\ell^2} \left[ -\hat{R}_1-(m^2)_1\frac{(3r_m^6+2r_c^6 - 3r_0^2 r_c^4 -2r_0^6)}{(r_m^6-r_0^6)}\right]
   \end{align}
where $t_{\text{I}}, t_{\text{II}}, b_{\text{I}}, b_{\text{II}}$ are integration constants fixed by the boundary conditions $\delta\alpha_1(r_0)=0$ and $\delta\alpha_1(r_m)=\delta a_1$ with $\delta\alpha$ and $\delta\alpha'$ continuous at $r_c$.

\begin{figure}[t]
\centering
\begin{subfigure}[t]{0.47\textwidth}
\includegraphics[width=\textwidth]{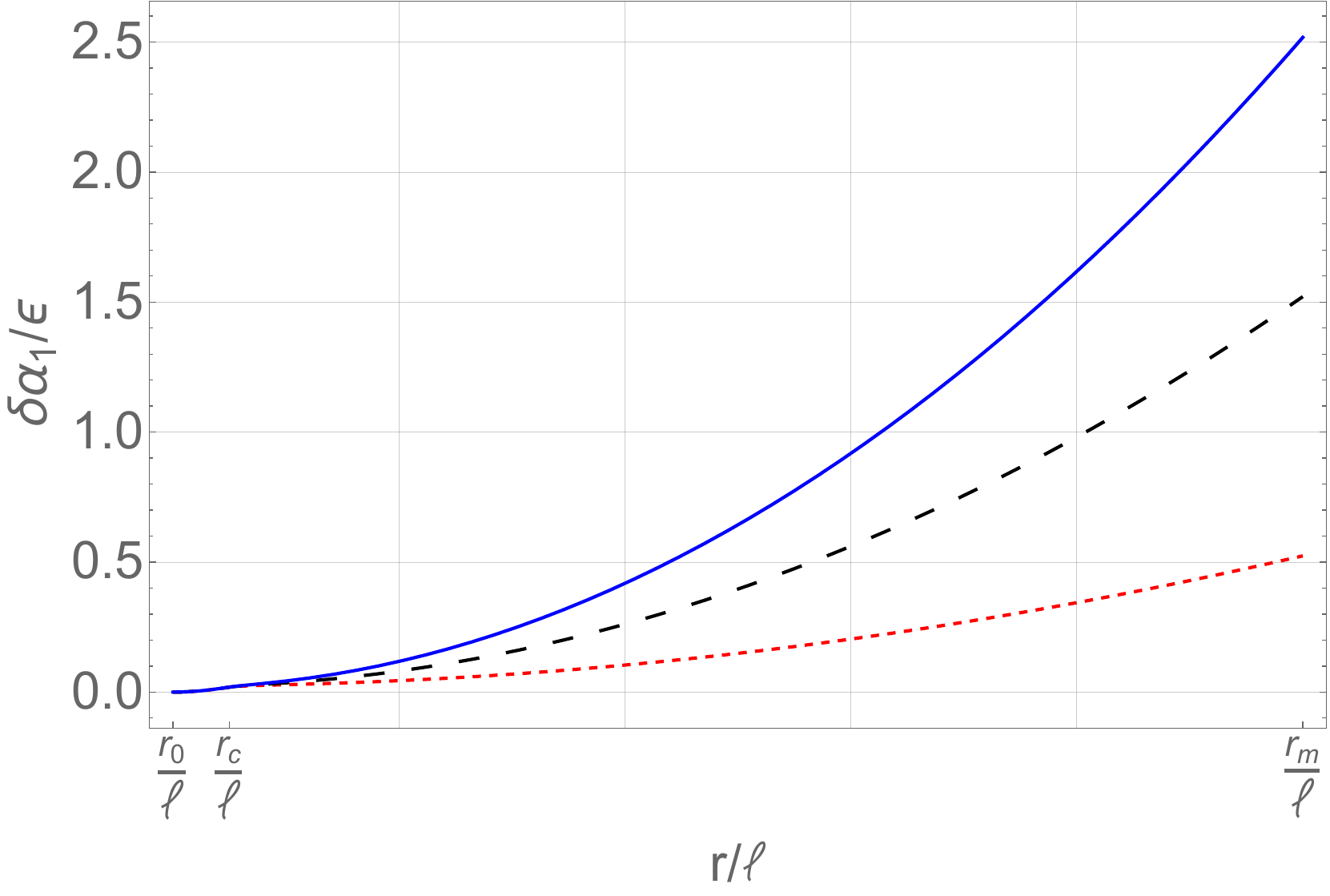}
\caption{$r_c=0.5\ell$}
\label{fig:paradep1rc05}
\end{subfigure}
\begin{subfigure}[t]{0.47\textwidth}
\includegraphics[width=\textwidth]{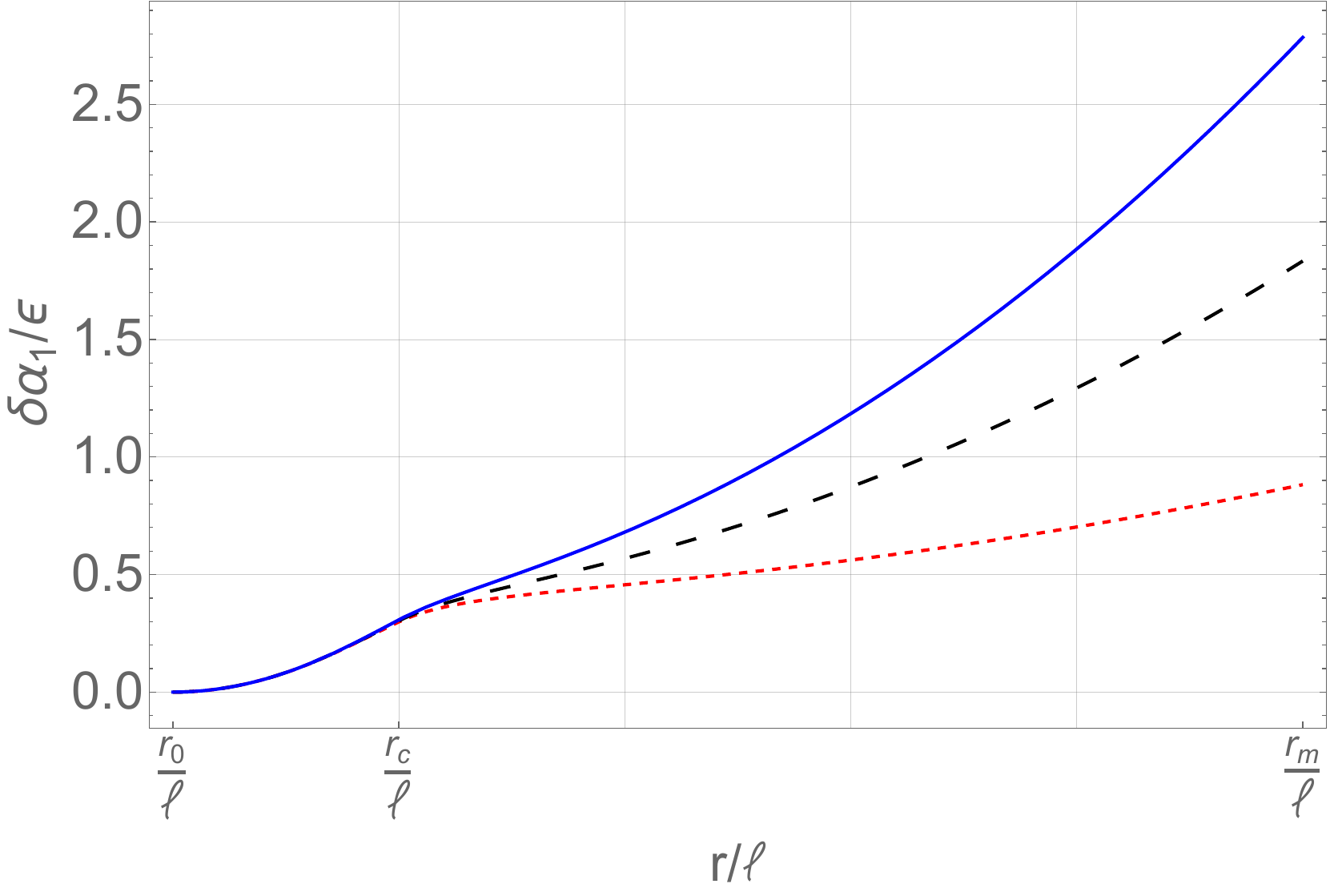}
\caption{$r_c=2\ell$}
\label{fig:paradep1rc2}
\end{subfigure}
\caption{$\delta\alpha_1(r)/\epsilon$ vs $r/\ell$ for two different values of $r_c$ (as labeled in subfigures) and three values $\delta a_1=0.1\epsilon$ (red dotted), $0.3\epsilon$ (black dashed), $0.5\epsilon$ (blue solid). The range is $r_0\leq r\leq r_m$ with $r_0=\ell/1000$ and $r_m=10\ell$.}
\label{fig:paradep1}
\end{figure}

The boundary conditions give us the mass squared
\begin{equation}
    (m^2)_1 =-\frac{\hat{R}_1}{3}-\frac{2 r_m^5}{(r_m^6-r_0^6)}\delta a_1 ,
\end{equation}
which is
\begin{equation}
    (m^2)_1\sim-\frac{\hat{R}_1}{3}-\frac{2}{r_m}\delta a_1 
\end{equation}
in the long throat limit.
Again, we see that the mass goes to a constant rather than scaling with the warp factor at the bottom of the throat.
In the large bulk volume limit, i.e. when $r_c \to r_0$, we find 
\begin{equation}
    (m^2)_1\sim-\frac{\hat{R}_1}{3}-\frac{2 r_m^5}{(r_m^6 - r_0^6)}\delta a_1 . 
\end{equation}
Notice that this reduces to the result for the mass (\ref{eq:MassLargeVolume}) in the large volume limit of section \ref{sec:largeVolume} in the case of a vanishing throat $r_0 \rightarrow 0$.

Figure \ref{fig:paradep1} illustrates the profile of the volume mode through the function $\delta\alpha_1(r)$ for two different values of $r_c$ (representing the stabilized value of the volume modulus) and three values for the boundary condition $\delta a_1$. 
For the purposes of the figure, we define the small parameter $\epsilon$ through $\hat R_1 \equiv \epsilon/\ell^2$. 
The main feature to note is that $\delta\alpha_1$ does not accumulate in the throat, even though its equation of motion determines the mass of the volume modulus. 
Instead, it grows toward the UV of the throat and into the bulk.
The figure emphasizes our point that the mass does not redshift for long throats but rather goes to a constant value determined by the gaugino condensate and stabilized background.
We note that the usual ``wavefunction" for the modulus is the function $\delta f(y)$ in the fluctuating warp factor, but we plot $\delta\alpha$ because it controls the mass eigenvalue.

\subsubsection{Parametric scaling case 2}

\begin{figure}[t]
\centering
\begin{subfigure}[t]{0.47\textwidth}
\includegraphics[width=\textwidth]{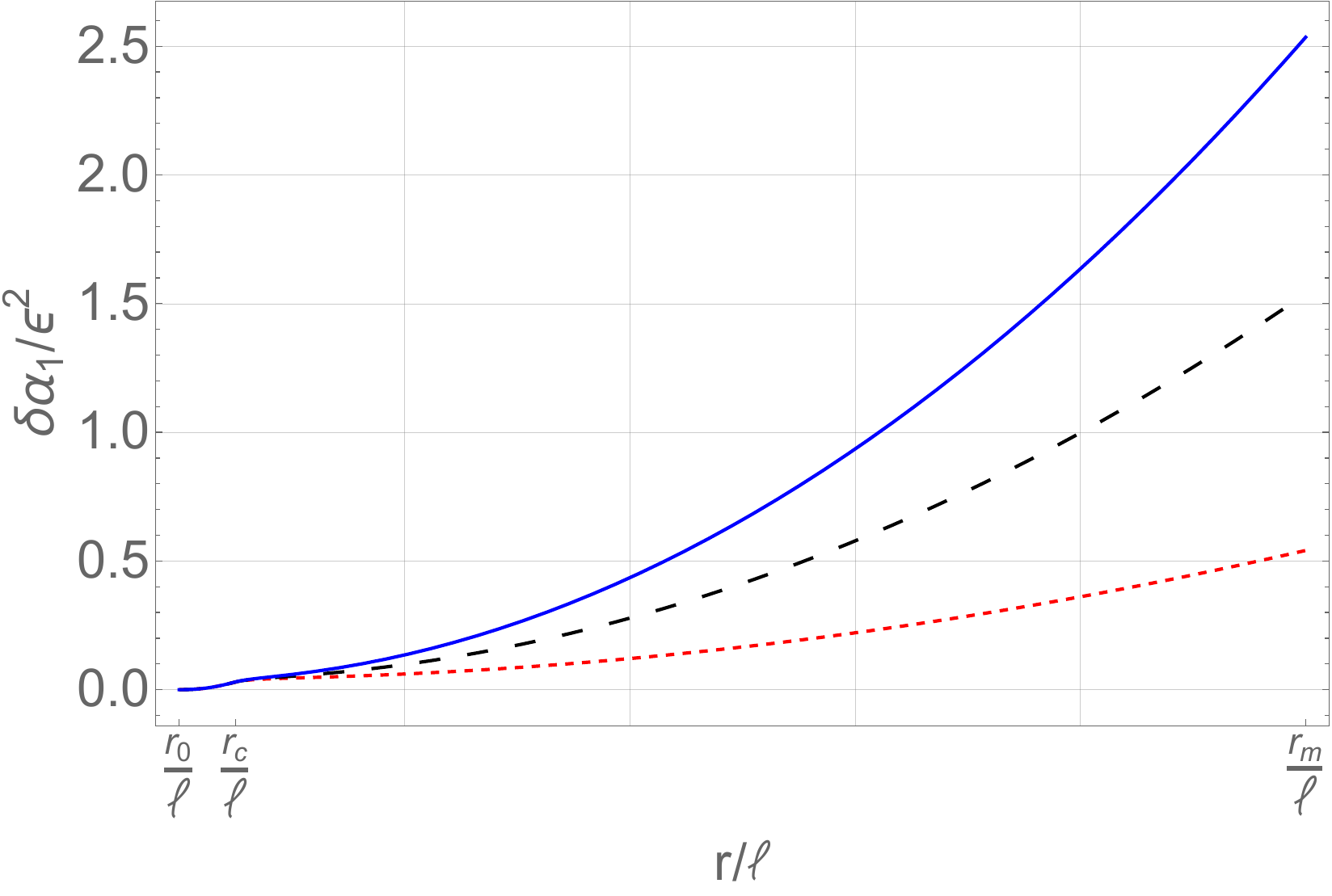}
\caption{$r_c=0.5\ell$}
\label{fig:paradep2rc05}
\end{subfigure}
\begin{subfigure}[t]{0.47\textwidth}
\includegraphics[width=\textwidth]{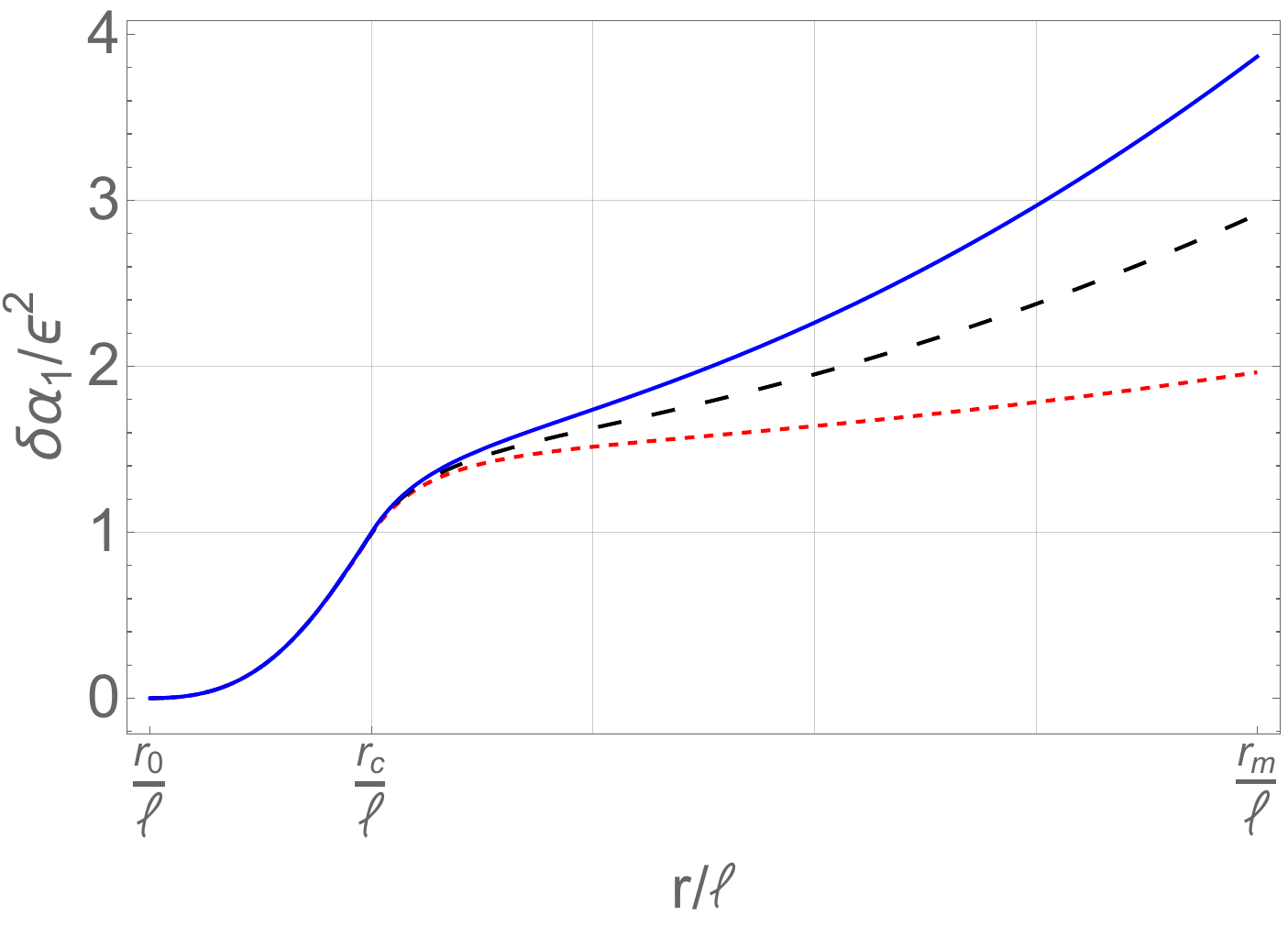}
\caption{$r_c=2\ell$}
\label{fig:paradep2rc2}
\end{subfigure}
\caption{$\delta\alpha_2(r)/\epsilon^2$ vs $r/\ell$ for two different values of $r_c$ (as labeled in subfigures) and three values $\delta a_2=0.1\epsilon^2$ (red dotted), $0.3\epsilon^2$ (black dashed), $0.5\epsilon^2$ (blue solid). The flux parameters are $g_1=3\epsilon \ell^{6-\delta}$ and $\delta=3/2$.  The range is $r_0\leq r\leq r_m$ with $r_0=\ell/1000$ and $r_m=10\ell$.}
\label{fig:paradep2}
\end{figure}

For the second case for the parametric dependence, we take the same boundary conditions $\delta\alpha'(r_0)=0$, $\delta\alpha'(r_m)=\delta a$.
After following the same rough arguments and including flux terms, we again estimate that the fluctuated boundary conditions should be $\delta a_2\approx \pm e^{2\Omega_0}a_2$.

\begin{figure}[t]
\centering
\begin{subfigure}[t]{0.47\textwidth}
\includegraphics[width=\textwidth]{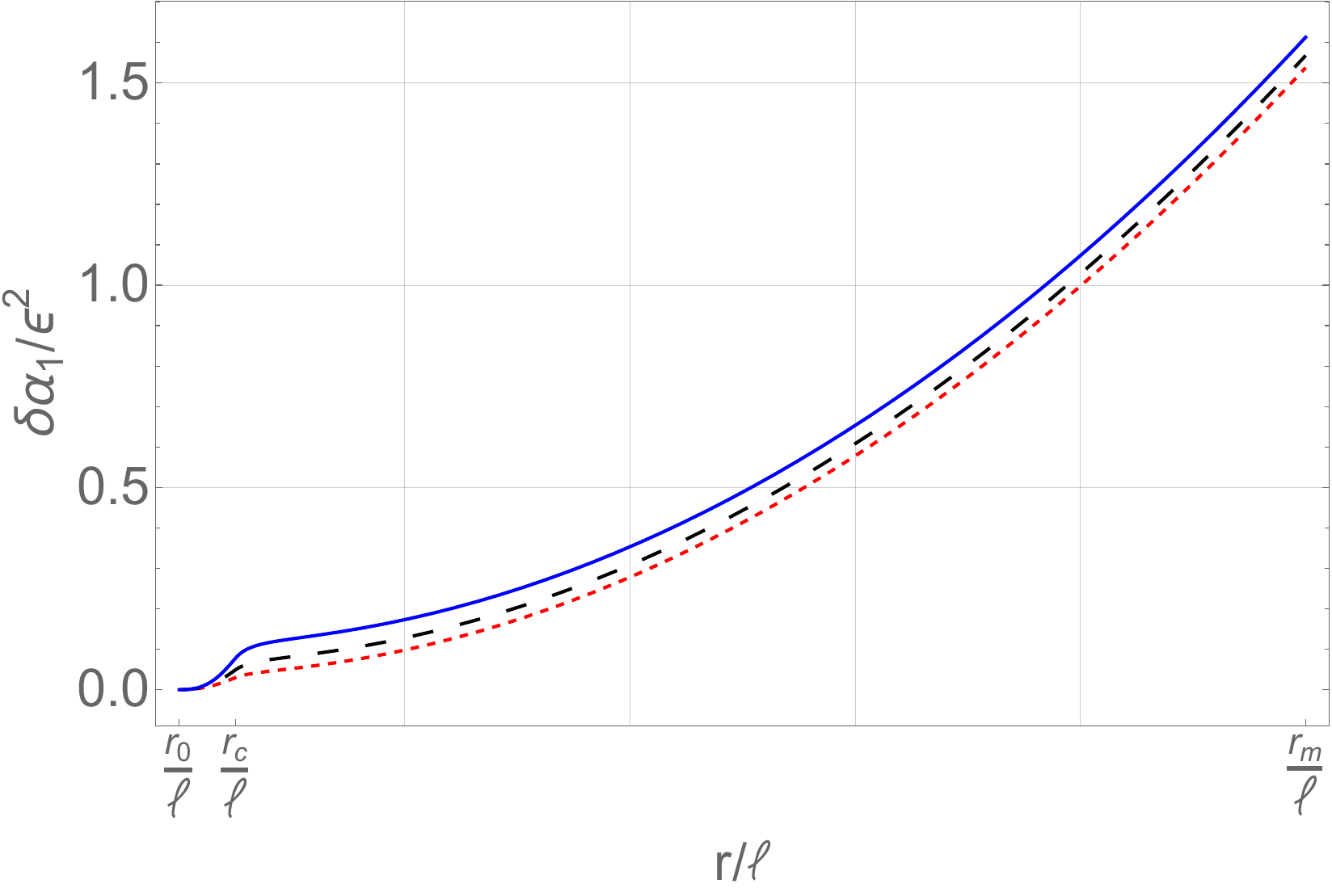}
\caption{$r_c=0.5\ell$}
\label{fig:paradep2rc05g}
\end{subfigure}
\begin{subfigure}[t]{0.47\textwidth}
\includegraphics[width=\textwidth]{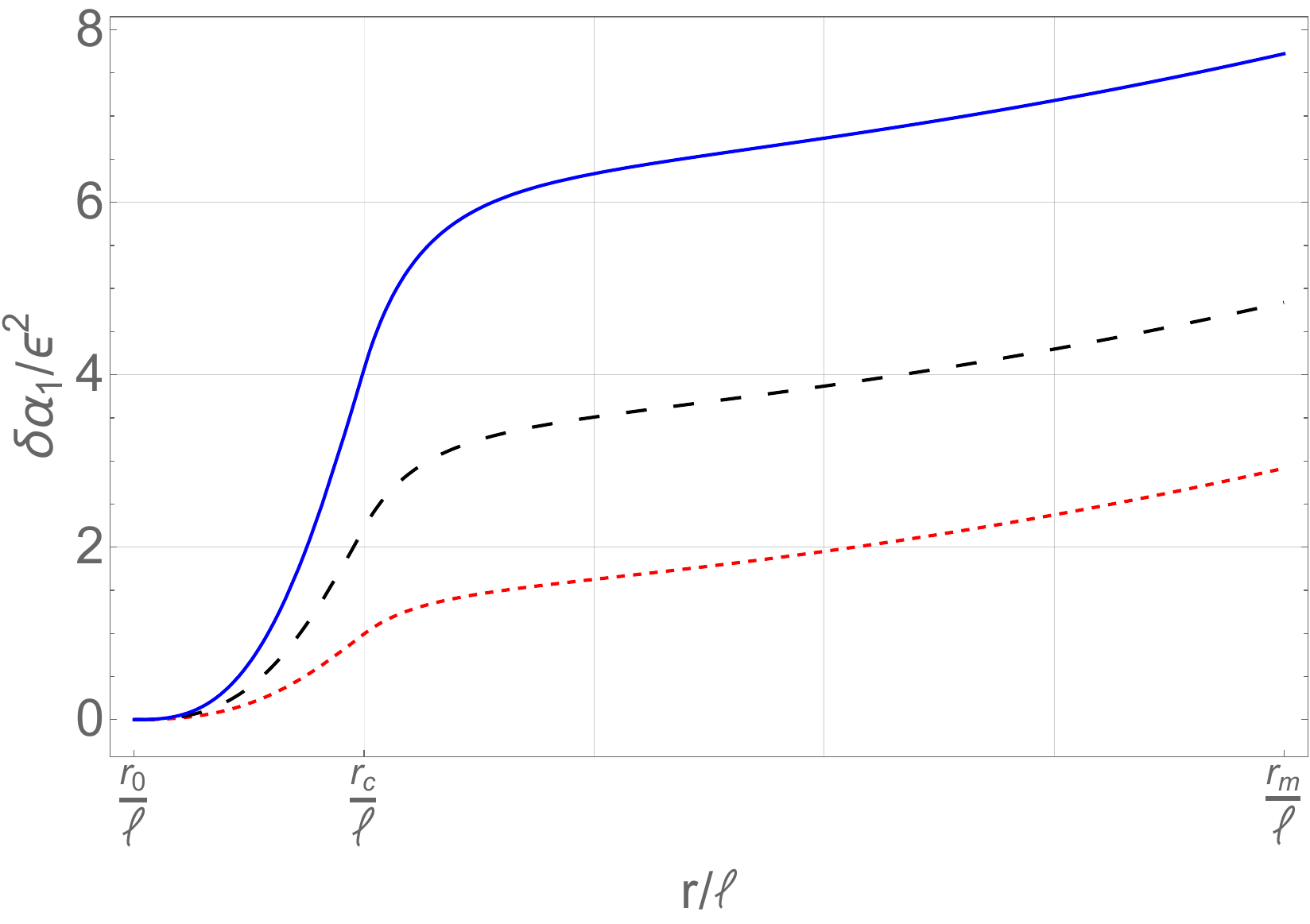}
\caption{$r_c=2\ell$}
\label{fig:paradep2rc2g}
\end{subfigure}
\caption{$\delta\alpha_2(r)/\epsilon^2$ vs $r/\ell$ for two different values of $r_c$ (as labeled in subfigures) and three values $g_1=3\epsilon\ell^6$ (red dotted), $0.3\epsilon^2$ (black dashed), $0.5\epsilon^2$ (blue solid). The flux parameters are $g_1=3\epsilon\ell^{6-\delta}$ (red dotted),  $3\epsilon\ell^{6-\delta}$ (black dashed), and $7\epsilon\ell^{6-\delta}$ (blue solid) and $\delta=3/2$ in all cases.  The range is $r_0\leq r\leq r_m$ with $r_0=\ell/1000$ and $r_m=10\ell$, and the boundary condition is $\delta a_2=0.3\epsilon^2$.}
\label{fig:paradep2g}
\end{figure}

In the throat,
\begin{align}
    \frac{\mathrm d^2 \delta \alpha_2}{\mathrm d r^2}+\frac{5}{r}\frac{\mathrm d\delta\alpha_2}{\mathrm d r} =& -\hat{R}_2-2(m^2)_2-e^{-2\Omega_0}e^{4A_0(y)}(m^2)_2 - 4(m^2)_2 \partial_r \delta K_0(y) \partial_r\left(e^{4A_0(y)}\right)\nonumber\\
&+g_1^2(\delta-2)^2 \frac{r^{2\delta-2}}{\ell^{12}} .
\end{align}
Meanwhile, in the bulk,
\begin{equation}
    \frac{\mathrm d^2 \delta \alpha_2}{\mathrm d r^2}+\frac{5}{r}\frac{\mathrm d\delta\alpha_2}{\mathrm d r}= -\hat{R}_2-2(m^2)_2-e^{-2\Omega_0}e^{4A_0(y)}(m^2)_2+g_1^2(\delta-2)^2 \frac{r_c^{2\delta-2}}{\ell^{12}} .
\end{equation}
We solve for $\delta\alpha_2$ the aforementioned boundary and matching conditions to determine the mass. 
The general formula for the mass is long and not particularly illuminating, so we present two limits.
First, in the long throat limit $r_0\to 0$, the mass is given by
\begin{equation}
    (m^2)_2\sim-\frac{\hat{R}_2}{3}-\frac{2}{r_m}\delta a_2 
    -\frac{g_1^2(\delta -2)^2 r_c^{2 \delta -2} \left(r_c^6(\delta-1)-r_m^6(\delta+2)\right)}{3\ell^{12}r_m^6(\delta+2)}.
\end{equation}
The astute reader will note that this case follows the same pattern as the others: the mass goes to a constant rather than scaling with the warp factor at the bottom of the throat.
In the large volume regime $r_c\to r_0$,  the mass reduces to 
\begin{equation}
    (m^2)_2\sim-\frac{\hat{R}_2}{3} -\frac{2 r_m^5}{(r_m^6-r_0^6)}\delta a_2 
    +\frac{g_1^2(\delta -2)^2 r_0^{2 \delta -2}}{3 \ell^{12}}\,.
\end{equation}
Again, this reduces to the large volume limit result (\ref{eq:MassLargeVolume}) as it approaches infinite volume $r_0 \rightarrow 0$. Interestingly, in this limit, the contributions from 
the flux scales with a power of the warp factor and becomes negligible. However, the overall mass goes to a constant.

Figures \ref{fig:paradep2} and \ref{fig:paradep2g} illustrate the profile of the modulus for this parametric case for two values of the stabilized volume modulus and varying boundary conditions and flux backgrounds. As in the other parametric dependence, we see that the profile is largest in the bulk and definitively does not accumulate in the throat, even for strong warping. For the purpose of plotting, we have defined $\hat R_2=\epsilon^2/\ell^2$.

\subsection{General lessons}\label{sec:lessons}

In the previous subsections, we analyzed the eigenvalue equations (\ref{eqn:mastereqnfirstorder}),(\ref{eqn:mastereqncase2}) for the mass of the volume modulus for several different internal space backgrounds, separately considering two possible parametric scalings of the background with the perturbation from the gaugino condensate.
For background geometries we considered a large volume/weakly warped limit, in which the internal space consists of a flat bulk, a strongly AdS warped throat, and a strongly warped throat attached to a flat bulk region which interpolates between the first two backgrounds.
In each case, solving the eigenvalue problem gives an expression for the mass of the volume modulus that can be organized into three contributions
\begin{equation}
m^2 \simeq -\frac{\hat R}{3} + ({\rm boundary\, conds}) + (G^-\ {\rm flux})\, .
\label{eq:MassSummaryCase}
\end{equation}
The first, $-\hat R/3$, is a universal contribution found in all backgrounds and scalings, and is positive for a 4-dimensional AdS minimum. We will say more about this term below.
The second contribution to (\ref{eq:MassSummaryCase}) comes from the boundary conditions on the wavefunction $\delta \alpha$, which themselves are set by the variation of the gaugino condensation local terms.
As argued in Section \ref{sec:largeVolume}, we expect these boundary condition terms to be $\sim {\mathcal O}(1) \hat R$.
The third contribution to the mass comes from background AISD $G^-$ flux sourced by the ${\mathcal O}(\epsilon)$ gaugino condensate (\ref{eq:GminusFlux}), and only occurs for parametric scaling of the form $\left(\hat R, \alpha, \tilde R\right)\sim {\mathcal O}(\epsilon^2)$ and $G^-\sim {\mathcal O}(\epsilon)$. 
Since the background $\hat R$ (\ref{eq:alphamaster2}) in this scaling case is determined by a combination of the local terms and the $G^-$ flux,
through a similar argument as in \ref{sec:largeVolume}
we expect the contribution of $G^-$ flux and boundary conditions to the mass (\ref{eq:MassSummaryCase}) to also be $\sim {\mathcal O}(1) \hat R$, consistent with conjectures such as \cite{Gautason:2015tig,Gautason:2018gln,Lust:2019zwm,Lust:2022lfc}.
Altogether, then, an analysis of the warped volume modulus in the presence of a gaugino condensate leads to an expectation that the mass of the volume modulus is $m^2 \simeq -{\mathcal O}(1)\hat R$.
We see that the throat and bulk background of Section \ref{sec:throatandbulk} interpolates nicely in the respective limits between the large volume limit of Section \ref{sec:largeVolume} and the warped throat background of Section \ref{sec:throat}.

Interestingly, 
there is a very weak dependence of the mass and wavefunction of the volume modulus on the warp factor, even for a strongly warped throat. This is in contrast to the general expectation that mass scales are warped down in strongly warped backgrounds \cite{Randall:1999ee,DeWolfeGiddings,Kofman:2005yz}.
For example, fluctuations of a 10-dimensional scalar field $\Phi$ in the warped background (\ref{eq:BackgroundMetric}) with some bulk mass $m_{10}$ satisfy the eigenvalue equation \cite{Burgess:2006mn} (see also \cite{Greene:2000gh})
\begin{equation}
    \left(\nabla^2_{10} - m_{10}^2\right) \Phi(x,y) = \left[e^{-2A} m^2 + e^{4A}\tilde\nabla^2 - m_{10}^2\right] \Phi(x,y) = 0\, ,
    \label{eq:10dScalar}
\end{equation}
where we wrote $\hat \nabla^2 \Phi = m^2 \Phi$ in terms of the 4-dimensional mass $m$.
For strongly warped throat backgrounds, zero mode solutions to (\ref{eq:10dScalar}) have eigenvalue masses that are warped down to the IR scale at the tip of the throat $m^2\sim m_{10}^2 e^{2A_{\rm tip}}$ \cite{Burgess:2006mn}.
As another example, consider fluctuations of the dilaton where there is a bulk flux-induced mass $m_f^2 \sim n_f^2/\alpha'$, in terms of flux quantum numbers $n_f$, from a GKP background.
In the presence of a strongly warped throat, the mass of the dilaton zero mode is warped down to the IR scale at the tip of the throat \cite{Frey:2006wv},
\begin{equation}
    m_{\rm dilaton}^2 \sim e^{2A_{\rm tip}}\, m_f^2 \, ,
\end{equation}
and the wavefunction peaks at the tip of the throat.
As another example of this effect, the masses of KK-gravitons in the presence of a strongly warped throat are warped down by the IR scale at the tip of the throat \cite{Randall:1999ee,DeWolfeGiddings,Firouzjahi:2005qs,Burgess:2006mn,ValeixoBento:2022qca},
\begin{equation}
    m_{\rm KK, grav} \sim \frac{e^{2A_{\rm tip}}}{R}\, .
\end{equation}
Preliminary analysis of the complex structure modulus of the warped deformed conifold \cite{Bena:2018fqc,Blumenhagen:2019qcg,Bena:2019sxm,Randall:2019ent,Dudas:2019pls,Lust:2022xoq} similarly finds that the mass of the zero mode appears to be suppressed by the warp factor at the tip of the throat.

A surprising result of our analysis is that the zero mode of the warped volume modulus is different, in that its mass and wavefunction are protected against warping corrections.
Indeed, in all cases there is a universal contribution $-\hat R/3$ to the volume modulus mass; for strongly warped backgrounds, this prevents a warping down of the mass to the IR scale of a warped throat.
Interestingly, this is a general feature of a scalar metric fluctuation that applies beyond the specific GKP background considered here.
Consider a 10-dimensional warped-product metric with general metric fluctuations parameterized by
\begin{align}
    ds^2 =& e^{2A(y)} \left[(1-2\psi(x,y))\hat g_{\mu\nu} + \hat \nabla_\mu \partial_\nu K(x,y)\right] dx^\mu dx^\nu + 2 e^{2A(y)} \partial_\mu B_m(x,y) dx^\mu dy^m \nonumber \\
    & e^{-2A(y)} \left[\tilde g_{mn} + 2 \phi_{mn}\right] dy^m dy^n
\end{align}
where $\hat g_{\mu\nu}$ is a maximally symmetric space with $\hat R [\hat g]$ constant.
As discussed in Section \ref{sec:intro}, these metric fluctuations organize themselves into the gauge-invariant combinations $\Phi_{mn}$ (\ref{eq:gaugeInvPhi}) and $\Psi$ (\ref{eq:gaugeInvPsi}) \cite{BreathingMode}.
The eigenvalue equation for the fluctuations arises from the internal $\delta G_{mn} - \kappa_{10}^2 \delta T_{mn}$ Einstein equations and includes the terms (after using the $(\mu\nu)$ constraint equations),
\begin{align}
    \delta G_{mn} - \kappa_{10}^2\delta T_{mn} \sim &\tilde{\mathcal L}_2[\Phi_{mn}] -e^{-4A} \left(\hat \Box + \hat R\right) \left(\Phi_{mn} + \frac{1}{2} \tilde g_{mn} \Phi^{\tilde m}_m\right) + \ldots \nonumber \\
    & \sim \tilde{\mathcal L}_2[\Phi_{mn}] -e^{-4A} \mu^2 \left(\Phi_{mn} + \frac{1}{2} \tilde g_{mn} \Phi^{\tilde m}_m\right) + \ldots =0
    \label{eq:GeneralEigenvalueEq}
\end{align}
where $\tilde{\mathcal L}_2[\Phi_{mn}]$ is a 2nd order operator on $\Phi_{mn}$ involving derivatives $\partial/\partial y^m$, the metric $\tilde g_{mn}$, and derivatives of the warp factor $\tilde\partial_m A$, $\mu^2 = m^2 + \hat R$ is the eigenvalue, and there are additional terms involving derivatives of the warp factor, internal curvature, and matter fields (including local sources) that depend on the details of the background that we will not need here.

Solving (\ref{eq:GeneralEigenvalueEq}) for some boundary conditions leads to an allowed value for the eigenvalue $\mu^2$; in general, this will depend on the background warping, the matter and its fluctuations, as well as other details related to the geometry. However, note that for any given eigenvalue $\mu^2$ the corresponding 4-dimensional mass is\footnote{\cite{Contaldi:2004hr} noticed this effect in the context of scalar metric perturbations of warped compactifications to 4-dimensional de-Sitter space, illustrating it with several explicit examples \cite{Gen:2000nu,Frolov:2003yi}.}
\begin{equation}
    m^2 = - \hat R + \mu^2 \, .
    \label{eq:GeneralMass}
\end{equation}
Conceptually, this explains the presence of the universal contribution $-\hat R/3$ to the volume modulus mass (the factor of $1/3$ is due to the specific form of the metric perturbation chosen as well as the use of other equations to obtain (\ref{eqn:mastereqnfirstorder}),(\ref{eqn:mastereqncase2})).
For AdS backgrounds, because of the form of the expression for the mass (\ref{eq:GeneralMass}), if the eigenvalue $\mu^2$ obtained from (\ref{eq:GeneralEigenvalueEq}) is exponentially warped down due to a strongly warped background as it is for other bulk modes \cite{Burgess:2006mn,Frey:2006wv,DeWolfeGiddings,Firouzjahi:2005qs,ValeixoBento:2022qca}, the 4-dimensional mass will nevertheless still be of the same order as the AdS curvature scale $\sim |\hat R|$, perhaps illustrating a higher-dimensional origin of the difficulty of separating the AdS scale and the mass scale of the lightest moduli \cite{Gautason:2018gln}.
For the warped volume modulus, we argued that $\mu^2\sim |\hat R|$ so that $m^2 \sim |\hat R|$ overall, but in principle it can be possible to achieve a separation of scales $m^2 \gg |\hat R|$ for a large-enough eigenvalue $\mu^2 \gg |\hat R|$, but this will require sources for $\mu^2$ in the Einstein equation that are parametrically decoupled from $\hat R$ (which does not appear to be the case for the volume modulus).

The universal contribution of the 4-dimensional curvature to the mass \eqref{eq:GeneralMass} bears a superficial resemblance to the Breitenlohner-Freedman (BF) bound on the mass of a scalar in AdS space. The origin of the two effects, however, are quite different.
The BF bound arises when recasting the Klein-Gordon equation for a scalar field in AdS space in Schrodinger-like form. The scalar field mass in this case is a free parameter, and the bound arises by demanding stable solutions.
In contrast, \eqref{eq:MassSummaryCase} is an expression determining the effective mass of the warped volume modulus in 4-dimensional AdS space as an eigenvalue equation. The contribution of $\hat R$ to the eigenvalue arises through the coupling of the 4-dimensional scalar curvature to the 10-dimensional volume modulus through the 10-dimensional equations of motion, rather than through the curved Laplacian of the 4-dimensional AdS space itself.
Thus, while the two effects share a surface similarity, the universal curvature contribution in our case reflects the dynamical coupling of the modulus to the 4-dimensional scalar curvature rather than a stability constraint.

\section{Discussion}\label{sec:discuss}

In this work, we made progress towards a 10-dimensional analysis of the
massive warped volume modulus stabilized by non-perturbative effects from gaugino condensation as in \cite{KKLT}.
As discussed in Section \ref{sec:WarpedVolModulus}, in order to construct a correct 4-dimensional effective theory of the volume modulus in warped backgrounds it is essential to have a thorough understanding of how the volume modulus degree of freedom appears in the 10-dimensional fields.
For example, an incorrect ansatz for the volume modulus in the metric can lead to an effective theory with different functional dependence, and the error is only detectable by examining the constraint equations that arise from the 10-dimensional equations of motion.
We extend the ansatz for the warped volume modulus of \cite{FTUD} for the 10-dimensional metric and fluxes to the background of \cite{KKLT} where the volume modulus is expected to be stabilized.

By solving the linearized supergravity equations of motion for the relevant fluctuations across several different geometries -- including a large-volume limit, a strongly warped throat, and a throat plus bulk system -- we arrived at a robust conclusion. Unlike other bulk modes in warped backgrounds studied previously \cite{DeWolfeGiddings,Burgess:2006mn,Frey:2006wv,Firouzjahi:2005qs,ValeixoBento:2022qca} in which the mass is warped down to the IR scale at the tip of the throat, we found that the mass of the warped volume modulus is largely independent of the warping, even for strongly warped throats.
This effect arises from a universal contribution to the mass-squared of the warped volume modulus appearing in the 10-dimensional equations of motion that is proportional to the negative of the background 4-dimensional Ricci curvature $\sim -\hat R$.
In an AdS vacuum where $\hat R$ is negative, this universal term provides a positive contribution to the mass-squared, bounding the mass from below from warp factor suppression.
For a 4-dimensional de Sitter or quasi-de Sitter background, the contribution of this universal term to the mass-squared is negative, and could play a role in destabilizing the volume modulus as in other models \cite{Gen:2000nu,Frolov:2003yi,Contaldi:2004hr}, perhaps along the lines discussed in \cite{Moritz:2017xto,Gautason:2018gln}.

For the warped volume modulus in the backgrounds of \cite{KKLT}, we argued that the other contributions to the mass are also $\sim {\mathcal O}(1) \hat R$, so that the overall mass of the warped volume modulus due to gaugino condensation is $m^2 \sim -{\mathcal O}(1) \hat R$.
One important consequence of this is that there is no clear parametric scale separation between the 4-dimensional curvature scale $\hat R$ and the volume modulus mass.
Indeed, our result $m^2 \sim -{\mathcal O}(1) \hat R$ from our 10-dimensional analysis is in mild disagreement with the mass of the volume modulus obtained from the KKLT effective field theory \cite{KKLT} $m^2_{\text{KKLT}}=-\mathcal{O}(100)\hat{R}$ (see Appendix \ref{app:KKLTMass}).
In the KKLT result, the ratio $m^2/\hat R$ can be written in terms of the stabilized volume modulus or alternatively by the tree-level superpotential (i.e., the flux quanta), which determines the stabilized value of the modulus.

It is important to ask whether this discrepancy between our results and the KKLT effective theory arises from a shortcoming in our 10-dimensional analysis or signals a genuine correction to the effective theory.
For example, we have considered only some of the constraints, so a modified ansatz for the volume modulus may be necessary; however, we have seen that changes to the ansatz to solve a different subset of the constraints make only minor changes to the master equation governing $\alpha(x,y)$.
Alternatively, the discrepancy could be due to incomplete knowledge of the background and how the gaugino condensate depends on the fully 10-dimensional warped volume modulus; these issues may also impact the construction of a complete ansatz for the fluctuating modulus.
More fundamentally, the tree-level supergravity with additional sources due to the gaugino condensate may not be the correct 10-dimensional theory to describe the background or fluctuations; \cite{Koerber:2007xk,Koerber:2008sx,Bena:2019mte,Grana:2020hyu,Grana:2022nyp} have advocated for modified supersymmetry variations in the presence of the condensate, which argues that nonperturbative (or $\alpha'$) corrections should be important in the equations of motion.
In contrast, perhaps the simplest possibility is that there is a slightly fine-tuned cancellation between flux and local contributions to the Ricci scalar on the background that does not occur for fluctuations around the background.
Whatever the case, if the KKLT effective theory is correct, some contribution to the effective potential must depend strongly on the fluctuating volume modulus in comparison to the scale of the potential itself.

Our results relate directly to one form of the AdS scale separation conjecture of the swampland program \cite{Gautason:2018gln}, which states that moduli in an effective field theory with AdS minimum have mass of order the AdS scale. 
If correct, the conjecture poses a challenge for constructions of de Sitter backgrounds in string theory that rely on ``uplifting'' AdS backgrounds, since the uplifting potential has a tendency to destabilize potentials without sufficiently curved minima. 
Our 10-dimensional analysis agrees broadly with this scale separation conjecture, though there are possible loopholes as outlined above.


\acknowledgments

We would like to acknowledge R.~Mahanta for useful discussions. The work of NA and ARF has been supported by the Natural Sciences and Engineering Research Council of Canada Discovery Grant program, grant 2020-00054, and by the University of Winnipeg.

\appendix
\section{Detailed equations of motion}
\label{app:calculation}
\subsection{10-Dimensional Einstein Equations} 

Based on the ansatz for the metric, the flux $G_{(3)}$, the five form and for the axiodilaton 
\begin{align}\label{eqn:LinearPerAnsatz}
ds^2&=e^{2\Omega(x)}e^{2A(x,y)}\hat{g}_{\mu\nu}(x)dx^{\mu}dx^{\nu}+2e^{2\Omega(x)}e^{2A(x,y)}\partial_\mu B_m(x,y)dx^\mu dy^m \nonumber\\
    &+ e^{-2A(x,y)}\tilde{g}_{mn}(y)dy^m dy^n\\
    \tilde{F}_{(5)} &= \tilde
    {\star}\tilde{d}e^{-4A(x,y)}-e^{2\Omega(x)}\hat{d}(\tilde{\star}\tilde{d}L(x,y))+e^{4\Omega(x)}\hat{\epsilon}\wedge \tilde{d}\left(e^{4A(x,y)}+\alpha(x,y)\right)\nonumber\\
    &-e^{-8A(x,y)}\tilde{\star}\tilde{d}\alpha(x,y)+e^{4\Omega}\hat\star\hat dB\wedge\tilde d\left(e^{4A(x,y)}+\alpha(x,y)\right)
    -e^{2\Omega(x)}\hat{\star}\hat{d}\tilde{d}L(x,y)\\
     G_{(3)}&=G^+_{(3)}(x,y) + G^-_{(3)}(x,y)\\
     \tau(y) &= \tau_0+\tau_1(x,y)+\cdots ,
\end{align}
we can determine the Einstein tensor and the stress energy tensor. 
Note that the (A)ISD condition is defined with respect to the unwarped metric $\tilde g_{mn}$, which does not fluctuate, and therefore the full (A)ISD flux $G_{(3)}^\pm$ separates into (A)ISD background and fluctuation components with respect to the unwarped background metric. That is, for our ansatz, decomposition into (A)ISD components commutes with extracting the first-order fluctuation.

\paragraph{Einstein Tensor:}
The Einstein tensor for the ansatz (\ref{eqn:LinearPerAnsatz}) is given as
\begin{align}
    G_{\mu\nu}&=\hat{G}_{\mu\nu}-\frac{1}{2}e^{4A}e^{2\Omega}\hat{g}_{\mu\nu}\tilde{R}-\hat{g}_{\mu\nu}\left(-2\hat{\nabla}^{\hat{2}}\Omega+4\hat{\nabla}^{\hat{2}} A+e^{4A}e^{2\Omega}\hat{\nabla}^{\hat{2}} \tilde{\nabla}^{\tilde{a}}B_a\right)\nonumber\\
    &\quad -2                       e^{4A}e^{2\Omega}\hat{g}_{\mu\nu}\left(\tilde{\nabla}^{\tilde{2}}A-2\tilde{\nabla}_m A\tilde{\nabla}^{\tilde{m}}A\right)
    +\left(-2\hat{\nabla}_\mu \hat{\nabla}_\nu\Omega+4\hat{\nabla}_\mu \hat{\nabla}_\nu A+e^{4A}e^{2\Omega}\hat{\nabla}_\mu \hat{\nabla}_\nu \tilde{\nabla}^{\tilde{a}}B_a\right)\nonumber\\
    G_{\mu m}&=-\frac{1}{2}\partial_\mu B_m \hat{R}-\frac{1}{2}e^{4A+2\Omega}\partial_\mu B_m \tilde{R}-2e^{4A}e^{2\Omega}\partial_\mu B_m\left(\tilde{\nabla}^{\tilde{2}}A-2\tilde{\nabla}_m A\tilde{\nabla}^{\tilde{m}}A\right)\nonumber\\
    &\quad +\frac{1}{2}e^{4A}e^{2\Omega}\left[\hat{\nabla}_\mu\tilde{\nabla}^{\tilde{c}}(\tilde{d}B)_{mc}+4\tilde{\nabla}^{\tilde{a}}A\hat{\nabla}_\mu(\tilde{d}B)_{ma}\right]-\frac{1}{2}e^{4A}\hat{\nabla}_{\mu}\tilde{\nabla}_m e^{-4A}\nonumber\\
    G_{mn}&=\tilde{G}_{mn}-\frac{1}{2}e^{-4A}e^{-2\Omega}\tilde{g}_{mn}\hat{R} + e^{-4A}e^{-2\Omega}\tilde{g}_{mn}(3\hat{\nabla}^{\hat{2}}\Omega-2\hat{\nabla}^{\hat{2}}A)-8\tilde{\nabla}_m A \tilde{\nabla}_n A + 4\tilde{g}_{mn}\tilde{\nabla}_a A\tilde{\nabla}^{\tilde{a}}A\nonumber\\
    &\quad +\left(\hat{\nabla}^{\hat{2}}\tilde{\nabla}_{(m}B_{n)}+4\hat{\nabla}^{\hat{2}}B_{(m}\tilde{\nabla}_{n)}A-\tilde{g}_{mn}\hat{\nabla}^{\hat{2}}\tilde{\nabla}^{\tilde{a}}B_a-2\tilde{g}_{mn}\tilde{\nabla}^{\tilde{a}}A\hat{\nabla}^{\hat{2}}B_a\right).
\end{align}
In the expressions for the Einstein tensor, all the explicit dependence on fluctuations has been reduced to first order. However, there is implicit fluctuation dependence via the warp factor and the Weyl factor. 

\paragraph{Stress-Energy Tensor:} We determine the stress-energy tensor through first order in fluctuations.

The five form anstaz given in (\ref{eqn:LinearPerAnsatz}) is self-dual. The stress energy tensor for $\tilde{F}_{(5)}$ is given as $ T_{MN}=(1/96)\tilde{F}_{MABCD}\tilde{F}_{N}{}^{ABCD}$ and its components can be written as 
\begin{align}
    T_{\mu\nu}&=-4\hat{g}_{\mu\nu}e^{2\Omega}e^{4A}\partial_m A \partial^{\tilde{m}}A-2\hat{g}_{\mu\nu}e^{2\Omega}\partial_m A \partial^{\tilde{m}}\alpha-\frac{1}{4}\hat{g}_{\mu\nu}e^{2\Omega}e^{-4A}\partial_m\alpha \partial^{\tilde{m}}\alpha\nonumber\\
    T_{\mu m}&=-4\partial_\mu B_m e^{2\Omega}e^{4A}\partial_b A \partial^{\tilde{b}}A-2\partial_\mu B_m e^{2\Omega}\partial_b A \partial^{\tilde{b}}\alpha-\frac{1}{4}\partial_\mu B_m e^{2\Omega}e^{-4A}\partial_b\alpha \partial^{\tilde{b}}\alpha\nonumber\\
    &\quad +\frac{1}{2}e^{2\Omega}\hat{\nabla}_{\mu}(\tilde{d}L)_{mb}\tilde{\nabla}^{\tilde{b}}\alpha+2e^{2\Omega}e^{4A}\hat{\nabla}_{\mu}(\tilde{d}L)_{mb}\tilde{\nabla}^{\tilde{b}}A\nonumber\\
    T_{mn}&=4\tilde{g}_{mn}\partial_j A \partial^{\tilde{j}}A-8\partial_m A \partial_n A + 2\tilde{g}_{mn}e^{-4A}\partial_j A \partial^{\tilde{j}}\alpha-2e^{-4A}\partial_m A \partial_n\alpha\nonumber\\
    &\quad -2e^{-4A}\partial_m \alpha \partial_n A -\frac{1}{2}e^{-8A}\partial_m \alpha \partial_n \alpha+\frac{1}{4}\tilde{g}_{mn}e^{-8A}\partial_j \alpha \partial^{\tilde{j}}\alpha.
\end{align}

Further, we can write the stress-energy tensor of the three form flux as $T_{MN}=(g_s/4)({G_{(M|}}^{PQ}\bar{G}_{|N)PQ}-g_{MN}|G_{(3)}|^2)$ and its components as 
\begin{align}
    T_{\mu\nu}&=-\frac{e^\phi}{4\times3!}e^{2\Omega}e^{8A}\hat{g}_{\mu\nu}\left(G^+_{mnr}\bar{G}^{+\widetilde{mnr}}+G^-_{mnr}\bar{G}^{-\widetilde{mnr}}\right)\nonumber\\
    T_{\mu m}&=-\frac{e^\phi}{4\times 3!}e^{2\Omega}e^{8A}\partial_\mu B_m(G^{+}_{abc}\bar{G}^{+\widetilde{abc}}+G^{-}_{abc}\bar{G}^{-\widetilde{abc}})\nonumber\\
    T_{mn}&=\frac{e^\phi}{4}e^{4A}\left[G_{(m|}^{+}{}^{\widetilde{pq}}\bar{G}^{-}_{|n)pq}+G_{(m|}^-{}^{\widetilde{pq}}\bar{G}^{+}_{|n)pq}\right]
\end{align}
where from the ISD/IASD property we can get the properties 
\begin{align}
        &{G^{+}_{(m|}}^{\widetilde{pq}}\bar{G}^{+}_{|n)pq}=\tilde{g}_{mn}G^+_{abc}\bar{G}^{+\widetilde{abc}}\\
        &{G^{-}_{(m|}}^{\widetilde{pq}}\bar{G}^{-}_{|n)pq}=\tilde{g}_{mn}G^-_{abc}\bar{G}^{-\widetilde{abc}}\\
        & G^-_{abc}\bar{G}^{+\widetilde{abc}}=0\\
        & G^+_{abc}\bar{G}^{-\widetilde{abc}}=0.
    \end{align}
This set of identities help us simplify the stress-energy tensor of the $G_{(3)}$ flux. 

Lastly, the stress-energy tensor for the axiodilaton is given as $T_{MN}=(1/2)e^{2\phi}\partial_{(M}\tau \partial_{N)}\bar{\tau}-(1/4)g_{MN}e^{2\phi}\partial_A \tau \partial^A \bar{\tau}$
and its components are 
\begin{align}
    T_{\mu\nu}&=-\frac{1}{4}e^{2\Omega}e^{4A}e^{2\phi}\hat{g}_{\mu\nu}\partial_a \tau \partial^{\tilde a}\bar \tau\nonumber\\
    T_{\mu m}&=\frac{1}{2}e^{2\phi}\partial_{(\mu}\tau\partial_{m)}\bar{\tau} -\frac{1}{4}e^{2\Omega}e^{4A}e^{2\phi}\partial_\mu B_m \partial_a \tau \partial^{\tilde a}\bar \tau\nonumber\\
    T_{mn}&=\frac{1}{2}e^{2\phi}\partial_{(m}\tau \partial_{n)}\bar \tau - \frac{1}{4}e^{2\phi}\tilde{g}_{mn}\partial_a\tau \partial^{\tilde a}\bar \tau.
\end{align}

\paragraph{Einstein Equations and Constraints:}
Since the AdS metric is Einstein ($\hat G_{\mu\nu}\propto \hat g_{\mu\nu}$), the $(\mu, \nu)$ Einstein equation separates into terms proportional to the metric and those proportional to the spacetime Hessian ($\hat{\nabla}_\mu \hat{\nabla}_\nu$) of the fluctuation, which are generally independent constraints. 
The latter, often called the ``off-diagonal'' constraint, is
\begin{equation}
    \left(-2\hat{\nabla}_\mu \hat{\nabla}_\nu\Omega+4\hat{\nabla}_\mu \hat{\nabla}_\nu A+e^{4A}e^{2\Omega}\hat{\nabla}_\mu \hat{\nabla}_\nu \tilde{\nabla}^{\tilde{a}}B_a\right)=0.
\end{equation}
Using this constraint we can simplify the $(\mu,\nu)$ Einstein equation to 
\begin{align}
    &\hat{G}_{\mu\nu}-\frac{1}{2}e^{4A}e^{2\Omega}\hat{g}_{\mu\nu}\tilde{R}-2e^{4A}e^{2\Omega}\hat{g}_{\mu\nu}\left(\tilde{\nabla}^{\tilde{2}}A-4\tilde{\nabla}_m A\tilde{\nabla}^{\tilde{m}}A\right)\nonumber\\
    &=-2\hat{g}_{\mu\nu}e^{2\Omega}\partial_m A \partial^{\tilde{m}}\alpha-\frac{1}{4}\hat{g}_{\mu\nu}e^{2\Omega}e^{-4A}\partial_m\alpha \partial^{\tilde{m}}\alpha\nonumber\\
    &-\frac{e^\phi}{4\times3!}e^{2\Omega}e^{8A}\hat{g}_{\mu\nu}\left( G^{+}_{abc}\bar{G}^{+\widetilde{abc}} + G^{-}_{abc}\bar{G}^{-\widetilde{abc}}\right)-\frac{1}{4}e^{2\Omega}e^{4A}e^{2\phi}\hat{g}_{\mu\nu}\partial_a \tau \partial^{\tilde a}\bar \tau.
\end{align}
For both the background and first-order fluctuations, this equation determines the $y$ dependence of the warp factor in terms of the other bulk fields.

Further, using the $(\mu,\nu)$ equation we can simplify the $(\mu,m)$ Einstein equation to
\begin{align}
    &-\frac{1}{2}e^{4A}\tilde{\nabla}_m\delta e^{-4A}-\frac{\hat{R}}{4}B_m + \frac{1}{2}e^{4A}e^{2\Omega}\tilde{\nabla}^{\tilde{c}}(\tilde{d}B)_{mc} + 2e^{4A}e^{2\Omega}(\tilde{d}B)_{mc}\tilde{\nabla}^{\tilde c}A\nonumber\\
    &-\frac{1}{2}e^{2\Omega}(\tilde{d}L)_{mc}\tilde{\nabla}^{\tilde{c}}\alpha - 2e^{2\Omega}e^{4A}(\tilde{d}L)_{mc}\tilde{\nabla}^{\tilde c}A-\frac{1}{4}e^{2\phi}\delta\tau\partial_m\bar{\tau}-\frac{1}{4}e^{2\phi}\delta\bar{\tau}\partial_m\tau=0
\end{align} 
This constraint is a relation between the exact and the co-exact part of the compensator and other fluctuations. 

The $(m,n)$ equation is 
\begin{align}\label{eq:mnEinstein}
&\tilde{G}_{mn}-\frac{1}{2}e^{-4A}e^{-2\Omega}\tilde{g}_{mn}\hat{R} + e^{-4A}e^{-2\Omega}\tilde{g}_{mn}(3\hat{\nabla}^{\hat{2}}\Omega-2\hat{\nabla}^{\hat{2}}A)\nonumber\\
&+\left(\hat{\nabla}^{\hat{2}}\tilde{\nabla}_{(m}B_{n)}+4\hat{\nabla}^{\hat{2}}B_{(m}\tilde{\nabla}_{n)}A-\tilde{g}_{mn}\hat{\nabla}^{\hat{2}}\tilde{\nabla}^{\tilde{a}}B_a-2\tilde{g}_{mn}\tilde{\nabla}^{\tilde{a}}A\hat{\nabla}^{\hat{2}}B_a\right)\nonumber\\
&=\frac{1}{4}\tilde{g}_{mn}e^{-8A}\partial_j \alpha \partial^{\tilde{j}}\alpha + 2\tilde{g}_{mn}e^{-4A}\partial_j A \partial^{\tilde{j}}\alpha-2e^{-4A}\partial_m A \partial_n\alpha-2e^{-4A}\partial_m \alpha \partial_n A-\frac{1}{2}e^{-8A}\partial_m \alpha \partial_n \alpha\nonumber\\
&+\frac{e^\phi}{4}e^{4A}\left[G_{(m|}^{+~~\widetilde{pq}}\bar{G}^{-}_{|n)pq} + G_{(m|}^{-~~\widetilde{pq}}\bar{G}^{+}_{|n)pq}\right]+\frac{1}{2}e^{2\phi}\partial_{(m}\tau \partial_{n)}\bar \tau - \frac{1}{4}e^{2\phi}\tilde{g}_{mn}\partial_a\tau \partial^{\tilde a}\bar \tau.
\end{align}
In principle, this gives the dynamical (spacetime) equation of motion for the fluctuation and therefore an eigenvalue equation for the mass. However, we will find a simpler combination of equations of motion to determine the mass.

\subsection{Form Equations of Motion}\label{app:form}
In this section, the $[i,j]$ component of a form has $i$ legs in the noncompact $x$ directions and $j$ legs in the compact $y$ directions.

\paragraph{$\tilde{F}_{(5)}$ equation of motion:} The equation of motion for the self-dual five form flux $\tilde{F}_{(5)}$ is 
\begin{equation}
    d\tilde{F}_{(5)}=\frac{i}{2}e^{\phi}G_{(3)}\wedge \bar{G}_{(3)}
\end{equation}
with $[0,6]$ component
\begin{align}
    \tilde{d}\tilde{\star}\tilde{d}e^{-4A}-\tilde{d}e^{-8A}\wedge \tilde{\star}\tilde{d}\alpha-e^{-8A}\tilde{d}\tilde{\star}\tilde{d}\alpha=\frac{e^{\phi}}{2}\left(G_{(3)}^+ \wedge \bar{G}^+_{(3)} + G_{(3)}^- \wedge \bar{G}^-_{(3)}\right).
\end{align}
The $[1,5]$ part of the equation of motion is
\begin{align}
    \tilde{\star}\tilde{d}e^{-4A}-e^{-8A}\tilde{\star}\tilde{d}\alpha+e^{2\Omega}\tilde{d}\tilde{\star}\tilde{d}L_1=0 ,
\end{align}
which simplifies to 
\begin{align}\label{eq:Lpoisson}
    -\partial_m e^{-4A}+e^{-8A}\partial_m \alpha + e^{2\Omega}\tilde{\nabla}^{\tilde{n}}(\tilde{d}L)_{mn}=0.
\end{align}
When combined with the $(\mu,m)$ Einstein equation, this gives us 
\begin{align}\label{eq:BminusL}
    &e^{4A}e^{2\Omega}\tilde{\nabla}^{\tilde c}\left[\tilde{d}(B-L)\right]_{mc}+ 4 e^{4A} e^{2\Omega}\tilde{\nabla}^{\tilde c}A \left[\tilde{d}(B-L)\right]_{mc} - e^{2\Omega}\tilde{\nabla}^{\tilde c}\alpha (\tilde{d}L)_{mc}\nonumber\\
    &= \frac{1}{2}B_m\hat{R}-e^{-4A}\partial_m \alpha-\frac{1}{2}e^{2\phi}\delta\tau\partial_m\bar{\tau}-\frac{1}{2}e^{2\phi}\delta\bar{\tau}\partial_m\tau.
\end{align}

\paragraph{$G_{(3)}$ equation of motion:} The equation of motion for the three form flux $G_{(3)}$ can be written as 
\begin{equation}
    d\star G_{(3)}=
    i\tilde{F}_{(5)}\wedge G_{(3)} -ie^{\phi}d\tau\wedge\star\textrm{Re}G_{(3)}.
\end{equation}
At the level of the background, this equation yields a $G_{(3)}$ as described by \cite{Baumann:2010sx} for the simplified warped throat. 
We wish to estimate the radial dependence of the fluctuations of $G_{(3)}$ with legs in all internal directions for the second case of parametric scaling in which $\hat R,m^2=\O(\epsilon^2)$.

We consider linear fluctuations of the background of the form 
\begin{equation}
    G_{(3)} = G_{(3)}^+ + G_{(3)}^- +c(x)\delta G_{(3)}^-(y)+ \hat d c(x) \delta A_{(2)}(y)
\end{equation}
where $\delta G_{(3)}^-$ is the same order in $\epsilon$ as $G_{(3)}^-$. 
Specifically, we ignore fluctuations in the ISD part of $G_{(3)}$ and the axiodilaton for our estimate.
The $(4,4)$ part of the $G_{(3)}$ equation of motion is sufficient to estimate $\delta G_{(3)}^-(y)$ in this approximation, so we do not need to consider the $\delta A_{(2)}$ part of the fluctuation.
Because the master equation involves $\delta G^-_{(3)}$ only in the second parametric scaling case, the terms at first order in the gaugino condensate and fluctuation become
\begin{align}
    \tilde{d}(e^{4A_0}\tilde{\star}\delta G_1^-)= i\tilde{d}e^{4A_0}\wedge \delta G_1^- +3ie^{8A_0}\tilde{d} G_1^-.
\end{align}
Schematically, this equation is roughly consistent with the estimate $\delta G_1^-\sim e^{4A_0}G_1^-$.


\section{Stabilized Breathing Mode of Freund-Rubin Compactification}\label{app:frbreathing}
In this section, we sketch the dimensional reduction of the stabilized breathing mode of FR compactification, emphasizing that it is possible to estimate the mass but not solve all the constraints without knowing the detailed background.
We start with the action 
\begin{align}
    S = \int d^{d+1+n}x\sqrt{-g_{d+1+n}}\, R_{d+1+n}+\frac{1}{2}\int(-1)^{n(d+1)}F_n\wedge\star F_n,
\end{align}
metric ansatz 
\begin{equation}
    ds^2=\hat{g}_{\mu\nu}dx^{\mu}dx^{\nu}+\tilde{g}_{mn}dy^m dy^n,
\end{equation}
and flux
\begin{equation}
    F_n =f \tilde{\epsilon}_n ,
\end{equation}
assuming that the $n$-form field threads the internal $n$ dimensions.  The external dimensions are $\textrm{AdS}_{d+1}$ with metric $\hat{g}_{\mu\nu}$. 
The internal dimensions are taken to be sphere $S^n$ with the metric $\tilde{g}_{mn}$. The stress energy for the $n$-form can be written as 
\begin{align}
    T_{MN}&=\frac{1}{2(n-1)!}F_{MA_1 \cdots A_{n-1}}F_{N}^{\;\;A_1 \cdots A_{n-1}}-\frac{1}{4n!}g_{MN}F_{A_1 \cdots A_n}F^{A_1 \cdots A_n}.
\end{align}
The background Einstein equations are
\begin{align}
    &\hat{R}_{\mu\nu}-\frac{1}{2}\hat{g}_{\mu\nu}(\hat{R}+\tilde{R})=-\frac{1}{4}\hat{g}_{\mu\nu}f^2\\
    &\tilde{R}_{mn}-\frac{1}{2}\tilde{g}_{mn}(\hat{R}+\tilde{R})=\frac{1}{4}\tilde{g}_{mn}f^2 .
\end{align}
For $d+1$ AdS spacetime, $\hat{R}_{\mu\nu}=-(d/L^2) \hat{g}_{\mu\nu}$, $L$ being the AdS length; for the $n$-sphere, $\tilde{R}_{mn}=((n-1)/\mathcal{R})\tilde{g}_{mn}$ where $\mathcal{R}$ is the radius of the sphere. Hence the background Einstein equations give two equations in the variables $L$ and $\mathcal{R}$ which can be solved as 
\begin{equation}
    L^2=\frac{2d(n+d-1)}{f^2(n-1)}\; ;\; \mathcal{R}^2=\frac{2(n-1)(n+d-1)}{df^2}. 
\end{equation}

Introducing a breathing mode fluctuation ansatz for the metric, we start with 
\begin{equation}
    ds^2=e^{\alpha u(x)}\hat{g}_{\mu\nu}dx^\mu dx^\nu + e^{2u(x)}\tilde{g}_{mn}dy^m dy^n
\end{equation}
where $\alpha=-2n/(d-1)$ is fixed by expressing the dimensionally reduced 4D action in Einstein frame. Linearizing the $(\mu,\nu)$ Einstein equation gives a constraint
\begin{equation}
    u(x)\left[-\frac{1}{2}(\alpha-2)\tilde{R}+\frac{1}{4}(\alpha-2n)f^2\right]=0.
\end{equation}
This equation is inconsistent unless 
\begin{equation}
    -\frac{1}{2}(\alpha-2)\tilde{R}+\frac{1}{4}(\alpha-2n)f^2=0
\end{equation}
which follows from the background solution. So we see that solving all the constraints requires knowledge of the background.

To find the mass of the fluctuation $u(x)$, we examine the $(m,n)$ Einstein equation, which reads, up to first order in $u(x)$
\begin{equation}
    m^2 \left(-1+n+\frac{d\alpha}{2}\right) + \frac{1}{2} (\alpha-2) \hat R = - \frac{1}{2} (n-1) f^2\, .
\end{equation}
As discussed in Section \ref{sec:lessons}, we see the presence of a universal $\hat R$ contribution to the mass.
Indeed, solving for the mass, we obtain
\begin{equation}
    m^2 = \frac{1}{2}\frac{-(\alpha-2)\hat R - (n-1) f^2}{-1+n+d\alpha/2} = -2 \frac{d}{d+1} \hat R\, ,
\end{equation}
where in the last equality we used the background solution to replace $f^2$ with a corresponding term proportional to $\hat R$, so that the final mass is $m^2 \sim {\mathcal O}(1) |\hat R|$, as expected. We emphasize, however, that estimating the mass in terms of both $\hat R$ and $f$ does not require detailed knowledge of the background solution, i.e., the relations between the curvatures and flux, even though solving the constraints does.

\section{KKLT Volume Modulus Mass}
\label{app:KKLTMass}

It is beneficial to briefly review the calculation that gives us the mass of the volume modulus in the 4D KKLT effective field theory as presented in \cite{KKLT}. The K{\"a}hler potential for the axion plus volume modulus complex scalar field is given by 
\begin{equation}
    K=-3\ln{[-i(\rho-\Bar{\rho})]}.
\end{equation}
This gives us the K{\"a}hler metric and its inverse as 
\begin{equation}
    K_{\rho\Bar{\rho}}=\frac{-3}{(\rho-\Bar{\rho})^2}; K^{\rho\Bar{\rho}}=-\frac{(\rho-\Bar{\rho})^2}{3}.
\end{equation}
 The superpotential is given as 
\begin{equation}
    W=W_0+Ae^{ia\rho}
\end{equation}
where $A, a$ and $W_0$ are real and do not depend on $\rho,\Bar{\rho}$. $W_0$ is the Gukov-Vafa-Witten superpotential and $Ae^{ia\rho}$ is the non-perturbative part of the superpotential. The scalar potential is hence given by 
\begin{equation}
    V=e^K(K^{\rho\Bar{\rho}}D_\rho W \overline{D_{\rho}W}-3|W|^2)
\end{equation}
This makes the Lagrangian 
\begin{equation}
    \mathcal{L}=-\frac{3}{(\rho-\Bar{\rho})^2}\partial_\mu \rho \partial^\mu \Bar{\rho}+\frac{A^2 a^2 i}{3(\rho-\Bar{\rho})}e^{ia(\rho-\Bar{\rho})}-\frac{Aa}{(\rho-\Bar{\rho})^2}\left[W_0(e^{ia\rho}+e^{-ia\Bar{\rho}})+2Ae^{ia(\rho-\Bar{\rho})}\right] .
\end{equation}
We then set $\rho=i\sigma$, where $\sigma$ is the volume modulus. We find 
\begin{equation}
    \mathcal{L}=\frac{3}{4\sigma^2}\partial_\mu \sigma \partial^\mu \sigma+\frac{A^2 a^2}{6\sigma}e^{-2a\sigma}+\frac{Aa}{2\sigma^2}\left[We^{-a\sigma}+Ae^{-2a\sigma}\right]
\end{equation}
The minimum of the potential lies at $\sigma=\sigma_{cr}$, which solves 
\begin{equation}
    W_0=-Ae^{-a\sigma_{cr}}\left(1+\frac{2}{3}a\sigma_{cr}\right) ,
\end{equation}
and the value of the potential at the minimum is 
\begin{equation}
    V_{min}=-\frac{a^2A^2e^{-2a\sigma_{cr}}}{6\sigma_{cr}} = \hat{R} ,
\end{equation}
where $\hat{R}$ is the Ricci scalar of the four dimensional AdS spacetime. In order to find the mass of the volume modulus, we turn the kinetic term into canonical form, by doing a variable change to $\tilde{\sigma}=\sqrt{\frac{3}{2}}\ln{\sigma}$ (also $\tilde{\sigma}_{cr}=\sqrt{\frac{3}{2}}\ln{\sigma_{cr}}$). We can expand our potential near $\tilde{\sigma}_{cr}$ to find the mass
\begin{equation}
    m^2=\frac{1}{9} a^2 A^2 e^{-2 a e^{\sqrt{\frac{2}{3}} \tilde{\sigma}_{cr}}-\sqrt{\frac{2}{3}} \tilde{\sigma}_{cr}} \left(a e^{\sqrt{\frac{2}{3}} \tilde{\sigma}_{cr}}+2\right) \left(2 a e^{\sqrt{\frac{2}{3}} \tilde{\sigma}_{cr}}+1\right) .
\end{equation}
Then
\begin{equation}
    m^2= -\frac{2}{3}\left(a e^{\sqrt{\frac{2}{3}} \tilde{\sigma}_{cr}}+2\right) \left(2 a e^{\sqrt{\frac{2}{3}} \tilde{\sigma}_{cr}}+1\right)\hat{R}.
\end{equation}
This tells us that, the mass of the volume modulus in the 4D KKLT EFT is proportional to the 4D Ricci scalar with proportionality generically  expected to be $\mathcal{O}(100)$ number. Because the modulus $\tilde\sigma_{cr}$ is evaluated at the potential minimum, the proportionality constant could be expressed either in terms of the volume modulus or discrete parameters of the model such as $W_0$.

\bibliographystyle{JHEP}
\bibliography{volume}

@article{KSThroat,
    author = "Klebanov, Igor R. and Strassler, Matthew J.",
    title = "{Supergravity and a confining gauge theory: Duality cascades and chi SB resolution of naked singularities}",
    eprint = "hep-th/0007191",
    archivePrefix = "arXiv",
    reportNumber = "IASSNS-HEP-00-56, PUPT-1944",
    doi = "10.1088/1126-6708/2000/08/052",
    journal = "JHEP",
    volume = "08",
    pages = "052",
    year = "2000"
}

@article{GKP,
    author = "Giddings, Steven B. and Kachru, Shamit and Polchinski, Joseph",
    title = "{Hierarchies from fluxes in string compactifications}",
    eprint = "hep-th/0105097",
    archivePrefix = "arXiv",
    reportNumber = "SLAC-PUB-8807, NSF-ITP-01-37, SU-ITP-01-16",
    doi = "10.1103/PhysRevD.66.106006",
    journal = "Phys. Rev. D",
    volume = "66",
    pages = "106006",
    year = "2002"
}

@article{DeWolfeGiddings,
    author = "DeWolfe, Oliver and Giddings, Steven B.",
    title = "{Scales and hierarchies in warped compactifications and brane worlds}",
    eprint = "hep-th/0208123",
    archivePrefix = "arXiv",
    reportNumber = "NSF-ITP-02-71, SU-ITP-02-27",
    doi = "10.1103/PhysRevD.67.066008",
    journal = "Phys. Rev. D",
    volume = "67",
    pages = "066008",
    year = "2003"
}

@article{KKLT,
      author         = "Kachru, Shamit and Kallosh, Renata and Linde, Andrei D.
                        and Trivedi, Sandip P.",
      title          = "{De Sitter vacua in string theory}",
      journal        = "Phys.Rev.",
      volume         = "D68",
      pages          = "046005",
      doi            = "10.1103/PhysRevD.68.046005",
      year           = "2003",
      eprint         = "hep-th/0301240",
      archivePrefix  = "arXiv",
      primaryClass   = "hep-th",
      reportNumber   = "SLAC-PUB-9630, SU-ITP-03-01, TIFR-TH-03-03",
      SLACcitation   = "%%CITATION = HEP-TH/0301240;%%",
}

@article{Giddings:2003zw,
    author = "Giddings, Steven B.",
    title = "{The Fate of four-dimensions}",
    eprint = "hep-th/0303031",
    archivePrefix = "arXiv",
    reportNumber = "MIFP-03-03",
    doi = "10.1103/PhysRevD.68.026006",
    journal = "Phys. Rev. D",
    volume = "68",
    pages = "026006",
    year = "2003"
}

@article{Hertzberg:2007wc,
    author = "Hertzberg, Mark P. and Kachru, Shamit and Taylor, Washington and Tegmark, Max",
    title = "{Inflationary Constraints on Type IIA String Theory}",
    eprint = "0711.2512",
    archivePrefix = "arXiv",
    primaryClass = "hep-th",
    reportNumber = "MIT-CTP-3905, SLAC-PUB-12999",
    doi = "10.1088/1126-6708/2007/12/095",
    journal = "JHEP",
    volume = "12",
    pages = "095",
    year = "2007"
}

@article{SimpledS,
    author = "Silverstein, Eva",
    title = "{Simple de Sitter Solutions}",
    eprint = "0712.1196",
    archivePrefix = "arXiv",
    primaryClass = "hep-th",
    reportNumber = "SLAC-PUB-13016, SITP-07-20",
    doi = "10.1103/PhysRevD.77.106006",
    journal = "Phys. Rev. D",
    volume = "77",
    pages = "106006",
    year = "2008"
}

@article{Haque:2008jz,
    author = "Haque, Sheikh Shajidul and Shiu, Gary and Underwood, Bret and Van Riet, Thomas",
    title = "{Minimal simple de Sitter solutions}",
    eprint = "0810.5328",
    archivePrefix = "arXiv",
    primaryClass = "hep-th",
    reportNumber = "SLAC-PUB-14712, MAD-TH-2008-13, SU-ITP-08-27, UG-FT-241-08, CAFPE-111-08",
    doi = "10.1103/PhysRevD.79.086005",
    journal = "Phys. Rev. D",
    volume = "79",
    pages = "086005",
    year = "2009"
}

@article{Danielsson:2009ff,
    author = "Danielsson, Ulf H. and Haque, Sheikh Shajidul and Shiu, Gary and Van Riet, Thomas",
    title = "{Towards Classical de Sitter Solutions in String Theory}",
    eprint = "0907.2041",
    archivePrefix = "arXiv",
    primaryClass = "hep-th",
    reportNumber = "UUITP-18-09, MAD-TH-09-06",
    doi = "10.1088/1126-6708/2009/09/114",
    journal = "JHEP",
    volume = "09",
    pages = "114",
    year = "2009"
}

@article{Wrase:2010ew,
    author = "Wrase, Timm and Zagermann, Marco",
    editor = "Anagnostopoulos, Konstantinos and Zoupanos, George",
    title = "{On Classical de Sitter Vacua in String Theory}",
    eprint = "1003.0029",
    archivePrefix = "arXiv",
    primaryClass = "hep-th",
    doi = "10.1002/prop.201000053",
    journal = "Fortsch. Phys.",
    volume = "58",
    pages = "906--910",
    year = "2010"
}

@article{VanRiet:2011yc,
    author = "Van Riet, Thomas",
    title = "{On classical de Sitter solutions in higher dimensions}",
    eprint = "1111.3154",
    archivePrefix = "arXiv",
    primaryClass = "hep-th",
    doi = "10.1088/0264-9381/29/5/055001",
    journal = "Class. Quant. Grav.",
    volume = "29",
    pages = "055001",
    year = "2012"
}

@article{KKLMMT,
      author         = "Kachru, Shamit and Kallosh, Renata and Linde, Andrei D.
                        and Maldacena, Juan Martin and McAllister, Liam P. and
                        others",
      title          = "{Towards inflation in string theory}",
      journal        = "JCAP",
      volume         = "0310",
      pages          = "013",
      doi            = "10.1088/1475-7516/2003/10/013",
      year           = "2003",
      eprint         = "hep-th/0308055",
      archivePrefix  = "arXiv",
      primaryClass   = "hep-th",
      reportNumber   = "SLAC-PUB-9669, SU-ITP-03-18, TIFR-TH-03-06",
      SLACcitation   = "%%CITATION = HEP-TH/0308055;%%",
}

@article{Frey:2002hf,
    author = "Frey, Andrew R. and Polchinski, Joseph",
    title = "{N=3 warped compactifications}",
    eprint = "hep-th/0201029",
    archivePrefix = "arXiv",
    reportNumber = "NSF-ITP-01-77",
    doi = "10.1103/PhysRevD.65.126009",
    journal = "Phys. Rev. D",
    volume = "65",
    pages = "126009",
    year = "2002"
}

@article{Giddings_Maharana,
      author         = "Giddings, Steven B. and Maharana, Anshuman",
      title          = "{Dynamics of warped compactifications and the shape of
                        the warped landscape}",
      journal        = "Phys.Rev.",
      volume         = "D73",
      pages          = "126003",
      doi            = "10.1103/PhysRevD.73.126003",
      year           = "2006",
      eprint         = "hep-th/0507158",
      archivePrefix  = "arXiv",
      primaryClass   = "hep-th",
      SLACcitation   = "%%CITATION = HEP-TH/0507158;%%",
}

@article{STUD,
      author         = "Shiu, Gary and Torroba, Gonzalo and Underwood, Bret and
                        Douglas, Michael R.",
      title          = "{Dynamics of Warped Flux Compactifications}",
      journal        = "JHEP",
      volume         = "0806",
      pages          = "024",
      doi            = "10.1088/1126-6708/2008/06/024",
      year           = "2008",
      eprint         = "0803.3068",
      archivePrefix  = "arXiv",
      primaryClass   = "hep-th",
      reportNumber   = "MAD-TH-08-05, RUNHETC-2008-03",
      SLACcitation   = "%%CITATION = ARXIV:0803.3068;%%",
}

@article{FTUD,
    author = "Frey, Andrew R. and Torroba, Gonzalo and Underwood, Bret and Douglas, Michael R.",
    title = "{The Universal Kahler Modulus in Warped Compactifications}",
    eprint = "0810.5768",
    archivePrefix = "arXiv",
    primaryClass = "hep-th",
    reportNumber = "RUNHETC-2008-19, NSF-KITP-08-133",
    doi = "10.1088/1126-6708/2009/01/036",
    journal = "JHEP",
    volume = "01",
    pages = "036",
    year = "2009"
}

@article{Frey:2006wv,
    author = "Frey, Andrew R. and Maharana, Anshuman",
    title = "{Warped spectroscopy: Localization of frozen bulk modes}",
    eprint = "hep-th/0603233",
    archivePrefix = "arXiv",
    reportNumber = "CALT-68-2591",
    doi = "10.1088/1126-6708/2006/08/021",
    journal = "JHEP",
    volume = "08",
    pages = "021",
    year = "2006"
}

@article{Burgess:2006mn,
    author = "Burgess, C. P. and Camara, Pablo G. and de Alwis, S. P. and Giddings, S. B. and Maharana, Anshuman and Quevedo, F. and Suruliz, Kerim",
    title = "{Warped Supersymmetry Breaking}",
    eprint = "hep-th/0610255",
    archivePrefix = "arXiv",
    reportNumber = "NSF-KITP-06-87, DAMTP-2006-78, CPHT-RR069-0906",
    doi = "10.1088/1126-6708/2008/04/053",
    journal = "JHEP",
    volume = "04",
    pages = "053",
    year = "2008"
}

@article{Greene:2000gh,
    author = "Greene, Brian R. and Schalm, Koenraad and Shiu, Gary",
    title = "{Warped compactifications in M and F theory}",
    eprint = "hep-th/0004103",
    archivePrefix = "arXiv",
    reportNumber = "CU-TP-971, NIKHEF-00-006, YITP-00-09",
    doi = "10.1016/S0550-3213(00)00400-4",
    journal = "Nucl. Phys. B",
    volume = "584",
    pages = "480--508",
    year = "2000"
}

@article{DT,
    author = "Douglas, Michael R. and Torroba, Gonzalo",
    title = "{Kinetic terms in warped compactifications}",
    eprint = "0805.3700",
    archivePrefix = "arXiv",
    primaryClass = "hep-th",
    reportNumber = "RUNHETC-2008-09",
    doi = "10.1088/1126-6708/2009/05/013",
    journal = "JHEP",
    volume = "05",
    pages = "013",
    year = "2009"
}

@article{Randall:2019ent,
    author = "Randall, Lisa",
    title = "{The Boundaries of KKLT}",
    eprint = "1912.06693",
    archivePrefix = "arXiv",
    primaryClass = "hep-th",
    doi = "10.1002/prop.201900105",
    journal = "Fortsch. Phys.",
    volume = "68",
    number = "3-4",
    pages = "1900105",
    year = "2020"
}

@article{Lust:2022xoq,
    author = {L\"ust, Severin and Randall, Lisa},
    title = "{Effective Theory of Warped Compactifications and the Implications for KKLT}",
    eprint = "2206.04708",
    archivePrefix = "arXiv",
    primaryClass = "hep-th",
    doi = "10.1002/prop.202200103",
    journal = "Fortsch. Phys.",
    volume = "70",
    number = "7-8",
    pages = "2200103",
    year = "2022"
}

@article{Firouzjahi:2005qs,
    author = "Firouzjahi, Hassan and Tye, S. -H. Henry",
    title = "{The Shape of gravity in a warped deformed conifold}",
    eprint = "hep-th/0512076",
    archivePrefix = "arXiv",
    doi = "10.1088/1126-6708/2006/01/136",
    journal = "JHEP",
    volume = "01",
    pages = "136",
    year = "2006"
}

@article{ValeixoBento:2022qca,
    author = "Valeixo Bento, Bruno and Chakraborty, Dibya and Parameswaran, Susha and Zavala, Ivonne",
    title = "{Gravity at the tip of the throat}",
    eprint = "2204.02086",
    archivePrefix = "arXiv",
    primaryClass = "hep-th",
    doi = "10.1007/JHEP09(2022)208",
    journal = "JHEP",
    volume = "09",
    pages = "208",
    year = "2022"
}

@article{BreathingMode,
      author         = "Underwood, Bret",
      title          = "{A Breathing Mode for Warped Compactifications}",
      journal        = "Class.Quant.Grav.",
      volume         = "28",
      pages          = "195013",
      doi            = "10.1088/0264-9381/28/19/195013",
      year           = "2011",
      eprint         = "1009.4200",
      archivePrefix  = "arXiv",
      primaryClass   = "hep-th",
      SLACcitation   = "%%CITATION = ARXIV:1009.4200;%%",
}

@article{Kachru:2019dvo,
    author = "Kachru, Shamit and Kim, Manki and Mcallister, Liam and Zimet, Max",
    title = "{de Sitter vacua from ten dimensions}",
    eprint = "1908.04788",
    archivePrefix = "arXiv",
    primaryClass = "hep-th",
    doi = "10.1007/JHEP12(2021)111",
    journal = "JHEP",
    volume = "12",
    pages = "111",
    year = "2021"
}

@article{1308.0323,
      author         = "Frey, Andrew R. and Roberts, James",
      title          = "{The Dimensional Reduction and K\"ahler Metric of Forms In
                        Flux and Warping}",
      journal        = "JHEP",
      volume         = "1310",
      pages          = "021",
      doi            = "10.1007/JHEP10(2013)021",
      year           = "2013",
      eprint         = "1308.0323",
      archivePrefix  = "arXiv",
      primaryClass   = "hep-th",
      SLACcitation   = "%%CITATION = ARXIV:1308.0323;%%",
}

@article{Baumann:2010sx,
    author = "Baumann, Daniel and Dymarsky, Anatoly and Kachru, Shamit and Klebanov, Igor R. and McAllister, Liam",
    title = "{D3-brane Potentials from Fluxes in AdS/CFT}",
    eprint = "1001.5028",
    archivePrefix = "arXiv",
    primaryClass = "hep-th",
    doi = "10.1007/JHEP06(2010)072",
    journal = "JHEP",
    volume = "06",
    pages = "072",
    year = "2010"
}

@article{Baumann:2008kq,
    author = "Baumann, Daniel and Dymarsky, Anatoly and Kachru, Shamit and Klebanov, Igor R. and McAllister, Liam",
    title = "{Holographic Systematics of D-brane Inflation}",
    eprint = "0808.2811",
    archivePrefix = "arXiv",
    primaryClass = "hep-th",
    reportNumber = "SLAC-PUB-13365, SU-ITP-08-19, ITEP-TH-17-08",
    doi = "10.1088/1126-6708/2009/03/093",
    journal = "JHEP",
    volume = "03",
    pages = "093",
    year = "2009"
}

@article{Hamada:2018qef,
    author = "Hamada, Yuta and Hebecker, Arthur and Shiu, Gary and Soler, Pablo",
    title = "{On brane gaugino condensates in 10d}",
    eprint = "1812.06097",
    archivePrefix = "arXiv",
    primaryClass = "hep-th",
    reportNumber = "CCTP-2018-15, ITCP-IPP 2018/11, MAD-TH-18-08",
    doi = "10.1007/JHEP04(2019)008",
    journal = "JHEP",
    volume = "04",
    pages = "008",
    year = "2019"
}

@article{Kallosh:2019oxv,
    author = "Kallosh, Renata",
    title = "{Gaugino Condensation and Geometry of the Perfect Square}",
    eprint = "1901.02023",
    archivePrefix = "arXiv",
    primaryClass = "hep-th",
    doi = "10.1103/PhysRevD.99.066003",
    journal = "Phys. Rev. D",
    volume = "99",
    number = "6",
    pages = "066003",
    year = "2019"
}

@article{Hamada:2019ack,
    author = "Hamada, Yuta and Hebecker, Arthur and Shiu, Gary and Soler, Pablo",
    title = "{Understanding KKLT from a 10d perspective}",
    eprint = "1902.01410",
    archivePrefix = "arXiv",
    primaryClass = "hep-th",
    reportNumber = "CCTP-2019-2, ITCP-IPP 2019/2, MAD-TH-19-02",
    doi = "10.1007/JHEP06(2019)019",
    journal = "JHEP",
    volume = "06",
    pages = "019",
    year = "2019"
}

@article{Gautason:2019jwq,
    author = "Gautason, F. F. and Van Hemelryck, V. and Van Riet, T. and Venken, V.",
    title = "{A 10d view on the KKLT AdS vacuum and uplifting}",
    eprint = "1902.01415",
    archivePrefix = "arXiv",
    primaryClass = "hep-th",
    doi = "10.1007/JHEP06(2020)074",
    journal = "JHEP",
    volume = "06",
    pages = "074",
    year = "2020"
}

@article{Kinoshita:2007uk,
    author = "Kinoshita, Shunichiro",
    title = "{New branch of Kaluza-Klein compactification}",
    eprint = "0710.0707",
    archivePrefix = "arXiv",
    primaryClass = "hep-th",
    reportNumber = "UTAP-586, RESCEU-86-07",
    doi = "10.1103/PhysRevD.76.124003",
    journal = "Phys. Rev. D",
    volume = "76",
    pages = "124003",
    year = "2007"
}

@article{Douglas:2009zn,
    author = "Douglas, Michael R.",
    title = "{Effective potential and warp factor dynamics}",
    eprint = "0911.3378",
    archivePrefix = "arXiv",
    primaryClass = "hep-th",
    doi = "10.1007/JHEP03(2010)071",
    journal = "JHEP",
    volume = "03",
    pages = "071",
    year = "2010"
}

@article{Smith:2024ejf,
    author = "Smith, George R. and Tennyson, David and Waldram, Daniel",
    title = "{All-orders moduli for type II flux backgrounds}",
    eprint = "2409.03847",
    archivePrefix = "arXiv",
    primaryClass = "hep-th",
    reportNumber = "Imperial/TP/24/DW/1; MI-HET-828",
    month = "9",
    year = "2024"
}

@article{Frey:2025rvf,
    author = "Frey, Andrew R. and Mahanta, Ratul",
    title = {{Dimensional Reduction and K\"ahler Metric for Metric Moduli in Imaginary Self-Dual Flux}},
    eprint = "2501.08623",
    archivePrefix = "arXiv",
    primaryClass = "hep-th",
    month = "1",
    year = "2025"
}

@article{Cownden:2016hpf,
    author = "Cownden, Brad and Frey, Andrew R. and Marsh, M. C. David and Underwood, Bret",
    title = "{Dimensional Reduction for D3-brane Moduli}",
    eprint = "1609.05904",
    archivePrefix = "arXiv",
    primaryClass = "hep-th",
    doi = "10.1007/JHEP12(2016)139",
    journal = "JHEP",
    volume = "12",
    pages = "139",
    year = "2016"
}

@article{Bena:2018fqc,
    author = {Bena, Iosif and Dudas, Emilian and Gra{\~n}a, Mariana and L{\"u}st, Severin},
    title = "{Uplifting Runaways}",
    eprint = "1809.06861",
    archivePrefix = "arXiv",
    primaryClass = "hep-th",
    reportNumber = "IPhT-T18/110, CPHT-RR080.082018",
    doi = "10.1002/prop.201800100",
    journal = "Fortsch. Phys.",
    volume = "67",
    number = "1-2",
    pages = "1800100",
    year = "2019"
}

@article{Blumenhagen:2019qcg,
    author = {Blumenhagen, Ralph and Kl{\"a}wer, Daniel and Schlechter, Lorenz},
    title = "{Swampland Variations on a Theme by KKLT}",
    eprint = "1902.07724",
    archivePrefix = "arXiv",
    primaryClass = "hep-th",
    reportNumber = "MPP-2019-21",
    doi = "10.1007/JHEP05(2019)152",
    journal = "JHEP",
    volume = "05",
    pages = "152",
    year = "2019"
}

@article{Bena:2019sxm,
    author = {Bena, Iosif and Buchel, Alex and L{\"u}st, Severin},
    title = "{Throat destabilization (for profit and for fun)}",
    eprint = "1910.08094",
    archivePrefix = "arXiv",
    primaryClass = "hep-th",
    reportNumber = "CPHT-RR058.102019",
    month = "10",
    year = "2019"
}

@article{Dudas:2019pls,
    author = {Dudas, Emilian and L{\"u}st, Severin},
    title = "{An update on moduli stabilization with antibrane uplift}",
    eprint = "1912.09948",
    archivePrefix = "arXiv",
    primaryClass = "hep-th",
    reportNumber = "CPHT-RR107.122019, IPhT-T19/167",
    doi = "10.1007/JHEP03(2021)107",
    journal = "JHEP",
    volume = "03",
    pages = "107",
    year = "2021"
}

@article{Contaldi:2004hr,
    author = "Contaldi, Carlo R. and Kofman, Lev and Peloso, Marco",
    title = "{Gravitational instability of de Sitter compactifications}",
    eprint = "hep-th/0403270",
    archivePrefix = "arXiv",
    reportNumber = "CITA-2004-04",
    doi = "10.1088/1475-7516/2004/08/007",
    journal = "JCAP",
    volume = "08",
    pages = "007",
    year = "2004"
}

@article{Gen:2000nu,
    author = "Gen, Uchida and Sasaki, Misao",
    title = "{Radion on the de Sitter brane}",
    eprint = "gr-qc/0011078",
    archivePrefix = "arXiv",
    reportNumber = "OU-TAP-150, UTAP-379",
    doi = "10.1143/PTP.105.591",
    journal = "Prog. Theor. Phys.",
    volume = "105",
    pages = "591--606",
    year = "2001"
}

@article{Frolov:2003yi,
    author = "Frolov, Andrei V. and Kofman, Lev",
    title = "{Can inflating brane worlds be stabilized?}",
    eprint = "hep-th/0309002",
    archivePrefix = "arXiv",
    reportNumber = "CITA-2003-29",
    doi = "10.1103/PhysRevD.69.044021",
    journal = "Phys. Rev. D",
    volume = "69",
    pages = "044021",
    year = "2004"
}

@article{Moritz:2017xto,
    author = "Moritz, Jakob and Retolaza, Ander and Westphal, Alexander",
    title = "{Toward de Sitter space from ten dimensions}",
    eprint = "1707.08678",
    archivePrefix = "arXiv",
    primaryClass = "hep-th",
    reportNumber = "DESY-17-112, DESY 17-112",
    doi = "10.1103/PhysRevD.97.046010",
    journal = "Phys. Rev. D",
    volume = "97",
    number = "4",
    pages = "046010",
    year = "2018"
}

@article{Gautason:2018gln,
    author = "Gautason, F. F. and Van Hemelryck, V. and Van Riet, T.",
    title = "{The Tension between 10D Supergravity and dS Uplifts}",
    eprint = "1810.08518",
    archivePrefix = "arXiv",
    primaryClass = "hep-th",
    doi = "10.1002/prop.201800091",
    journal = "Fortsch. Phys.",
    volume = "67",
    number = "1-2",
    pages = "1800091",
    year = "2019"
}

@article{Kallosh:2019axr,
    author = "Kallosh, Renata and Linde, Andrei and McDonough, Evan and Scalisi, Marco",
    title = "{dS Vacua and the Swampland}",
    eprint = "1901.02022",
    archivePrefix = "arXiv",
    primaryClass = "hep-th",
    doi = "10.1007/JHEP03(2019)134",
    journal = "JHEP",
    volume = "03",
    pages = "134",
    year = "2019"
}

@article{Carta:2019rhx,
    author = "Carta, Federico and Moritz, Jakob and Westphal, Alexander",
    title = "{Gaugino condensation and small uplifts in KKLT}",
    eprint = "1902.01412",
    archivePrefix = "arXiv",
    primaryClass = "hep-th",
    reportNumber = "DESY 19-012, DESY-19-012",
    doi = "10.1007/JHEP08(2019)141",
    journal = "JHEP",
    volume = "08",
    pages = "141",
    year = "2019"
}

@article{Bena:2019mte,
    author = "Bena, Iosif and Gra{\~n}a, Mariana and Kovensky, Nicolas and Retolaza, Ander",
    title = {{K{\"a}hler moduli stabilization from ten dimensions}},
    eprint = "1908.01785",
    archivePrefix = "arXiv",
    primaryClass = "hep-th",
    doi = "10.1007/JHEP10(2019)200",
    journal = "JHEP",
    volume = "10",
    pages = "200",
    year = "2019"
}

@article{Koerber:2008sx,
    author = "Koerber, Paul and Martucci, Luca",
    editor = "Lledo, Maria A.",
    title = "{Warped generalized geometry compactifications, effective theories and non-perturbative effects}",
    eprint = "0803.3149",
    archivePrefix = "arXiv",
    primaryClass = "hep-th",
    reportNumber = "MPP-2008-21, LMU-ASC-15-08",
    doi = "10.1002/prop.200810552",
    journal = "Fortsch. Phys.",
    volume = "56",
    pages = "862--868",
    year = "2008"
}

@article{Koerber:2007xk,
    author = "Koerber, Paul and Martucci, Luca",
    title = "{From ten to four and back again: How to generalize the geometry}",
    eprint = "0707.1038",
    archivePrefix = "arXiv",
    primaryClass = "hep-th",
    reportNumber = "MPP-2007-92, KUL-TF-07-12",
    doi = "10.1088/1126-6708/2007/08/059",
    journal = "JHEP",
    volume = "08",
    pages = "059",
    year = "2007"
}

@article{Grana:2020hyu,
    author = "Gra{\~n}a, Mariana and Kovensky, Nicolas and Retolaza, Ander",
    title = "{Gaugino mass term for D-branes and Generalized Complex Geometry}",
    eprint = "2002.01481",
    archivePrefix = "arXiv",
    primaryClass = "hep-th",
    reportNumber = "IPhT-t20/018",
    doi = "10.1007/JHEP06(2020)047",
    journal = "JHEP",
    volume = "06",
    pages = "047",
    year = "2020"
}

@article{Hamada:2021ryq,
    author = "Hamada, Yuta and Hebecker, Arthur and Shiu, Gary and Soler, Pablo",
    title = "{Completing the D7-brane local gaugino action}",
    eprint = "2105.11467",
    archivePrefix = "arXiv",
    primaryClass = "hep-th",
    reportNumber = "CTPU-PTC-21-23, MAD-TH-21-01",
    doi = "10.1007/JHEP11(2021)033",
    journal = "JHEP",
    volume = "11",
    pages = "033",
    year = "2021"
}

@article{Grana:2022nyp,
    author = "Gra{\~n}a, Mariana and Kovensky, Nicolas and Toulikas, Dimitrios",
    title = "{Smearing and unsmearing KKLT AdS vacua}",
    eprint = "2212.05074",
    archivePrefix = "arXiv",
    primaryClass = "hep-th",
    doi = "10.1007/JHEP03(2023)015",
    journal = "JHEP",
    volume = "03",
    pages = "015",
    year = "2023"
}

@article{Csaki:2000zn,
    author = "Csaki, Csaba and Graesser, Michael L. and Kribs, Graham D.",
    title = "{Radion dynamics and electroweak physics}",
    eprint = "hep-th/0008151",
    archivePrefix = "arXiv",
    reportNumber = "SCIPP-00-27",
    doi = "10.1103/PhysRevD.63.065002",
    journal = "Phys. Rev. D",
    volume = "63",
    pages = "065002",
    year = "2001"
}

@article{Randall:1999ee,
    author = "Randall, Lisa and Sundrum, Raman",
    title = "{A Large mass hierarchy from a small extra dimension}",
    eprint = "hep-ph/9905221",
    archivePrefix = "arXiv",
    reportNumber = "MIT-CTP-2860, PUPT-1860, BUHEP-99-9",
    doi = "10.1103/PhysRevLett.83.3370",
    journal = "Phys. Rev. Lett.",
    volume = "83",
    pages = "3370--3373",
    year = "1999"
}

@article{Kofman:2005yz,
    author = "Kofman, Lev and Yi, Piljin",
    title = "{Reheating the universe after string theory inflation}",
    eprint = "hep-th/0507257",
    archivePrefix = "arXiv",
    reportNumber = "KIAS-P05037",
    doi = "10.1103/PhysRevD.72.106001",
    journal = "Phys. Rev. D",
    volume = "72",
    pages = "106001",
    year = "2005"
}

@article{Gautason:2015tig,
    author = "Gautason, F. F. and Schillo, M. and Van Riet, T. and Williams, M.",
    title = "{Remarks on scale separation in flux vacua}",
    eprint = "1512.00457",
    archivePrefix = "arXiv",
    primaryClass = "hep-th",
    doi = "10.1007/JHEP03(2016)061",
    journal = "JHEP",
    volume = "03",
    pages = "061",
    year = "2016"
}

@article{Lust:2019zwm,
    author = {L{\"u}st, Dieter and Palti, Eran and Vafa, Cumrun},
    title = "{AdS and the Swampland}",
    eprint = "1906.05225",
    archivePrefix = "arXiv",
    primaryClass = "hep-th",
    doi = "10.1016/j.physletb.2019.134867",
    journal = "Phys. Lett. B",
    volume = "797",
    pages = "134867",
    year = "2019"
}

@article{Lust:2022lfc,
    author = {L{\"u}st, Severin and Vafa, Cumrun and Wiesner, Max and Xu, Kai},
    title = "{Holography and the KKLT scenario}",
    eprint = "2204.07171",
    archivePrefix = "arXiv",
    primaryClass = "hep-th",
    doi = "10.1007/JHEP10(2022)188",
    journal = "JHEP",
    volume = "10",
    pages = "188",
    year = "2022"
}

\end{document}